\documentclass[sigconf]{acmart}

\newcommand{\yujin}[1]{\textcolor{black}{#1}}
\newcommand{\jieshan}[1]{\textcolor{black}{#1}}
\newcommand{\jason}[1]{\textcolor{orange}{#1}}
\newcommand{\ccy}[1]{\textcolor{red}{#1}}
\newcommand{\chen}[1]{\textcolor{red}{#1}}
\newcommand{\zc}[1]{\textcolor{pink}{\textbf{ZC:} #1}}

\AtBeginDocument{%
  }
    
\usepackage{svg}
\usepackage{outlines}
\usepackage{subcaption} 
\usepackage{parskip}
\usepackage{float}
\usepackage{hyperref}
\usepackage{multirow} 
\usepackage{multicol}
\usepackage{graphicx} 
\usepackage{bookmark}
\bookmarksetup{depth=3}
\usepackage{xfp}
\usepackage[toc,page]{appendix}
\usepackage{enumitem}
\usepackage{colortbl}
\usepackage{graphics}
\usepackage{graphicx}
\usepackage{keyval}
\usepackage{trig}
\usepackage{array}
\usepackage{vcell}
\usepackage{color}

\newcommand \nApp{176}

\newcommand \nModelUsage{13}

\newcommand \nModelUsageTotal{255}
\newcommand \nAppCollected{62,822}

\renewcommand{\refname}{REFERENCE}

\newcommand \nIntPatternTotal{759}

\newcommand \nIntPatternGF{444}
\newcommand \nIntPatternFGF{214}

\newcommand \nIntPatternTextualFeedback{257}

\newcommand \nTextSuggestion{17}

\newcommand \AFnPatternGF{188}
\newcommand \AFnTextualFeedback{177}
\newcommand \AFnGraphicalFeedback{111}
\newcommand \AFnHapticNAuditory{36}

\newcommand \AFnIntPatternFGF{142}
\newcommand \AFFineGrainednGraphicalFeedback{132}

\newcommand \AFnIntPatternMF{97}

\newcommand \AFImpFeedback{78}
\newcommand \AFNoFeedback{22}

\setcopyright{none}
\renewcommand\footnotetextcopyrightpermission[1]{} 

\begin{document}

\title{Towards Real Smart Apps: Investigating Human-AI Interaction Patterns in Mobile On-Device AI Apps}

\author{Jason Ching Yuen Siu}
\email{csiu0002@student.monash.edu}
\affiliation{%
  \institution{Monash University}
  \country{Australia}
}

\author{Jieshan Chen}
\authornote{Corresponding Author.}
\orcid{0000-0002-2700-7478}
\email{Jieshan.Chen@data61.csiro.au}
\affiliation{%
  \institution{CSIRO's Data61}
  \country{Australia}
}

\author{Yujin Huang}
\email{yujin.huang@monash.edu}
\affiliation{%
  \institution{Monash University}
  \country{Australia}
}

\author{Zhenchang Xing}
\email{Zhenchang.Xing@data61.csiro.au}
\authornote{Also with Australian National University.}
\orcid{0000-0001-7663-1421}
\affiliation{%
  \institution{CSIRO's Data61}
  \country{Australia}
}

\author{Chunyang Chen}
\email{chunyang.chen@monash.edu}
\affiliation{%
  \institution{Monash University}
  \country{Australia}
}

\renewcommand{\shortauthors}{Siu et al.}

\begin{abstract}
With the emergence of deep learning techniques, smartphone apps are now embedded on-device AI features for enabling advanced tasks like speech translation, to attract users and increase market competitiveness. A good interaction design is important to make an AI feature usable and understandable. However, AI features have their unique challenges like sensitiveness to the input, dynamic behaviours and output uncertainty. Existing guidelines and tools either do not cover AI features or consider mobile apps which are confirmed by our informal interview with professional designers. To address these issues, we conducted the first empirical study to explore user-AI-interaction in mobile apps. We aim to understand the status of on-device AI usage by investigating 176 AI apps from 62,822 apps. We identified 255 AI features and summarized 759 implementations into three primary interaction pattern types. We further implemented our findings into a multi-faceted search-enabled gallery. The results of the user study demonstrate the usefulness of our findings.
\end{abstract}


\begin{teaserfigure}
	\centering
	\includegraphics[width=0.9\textwidth]{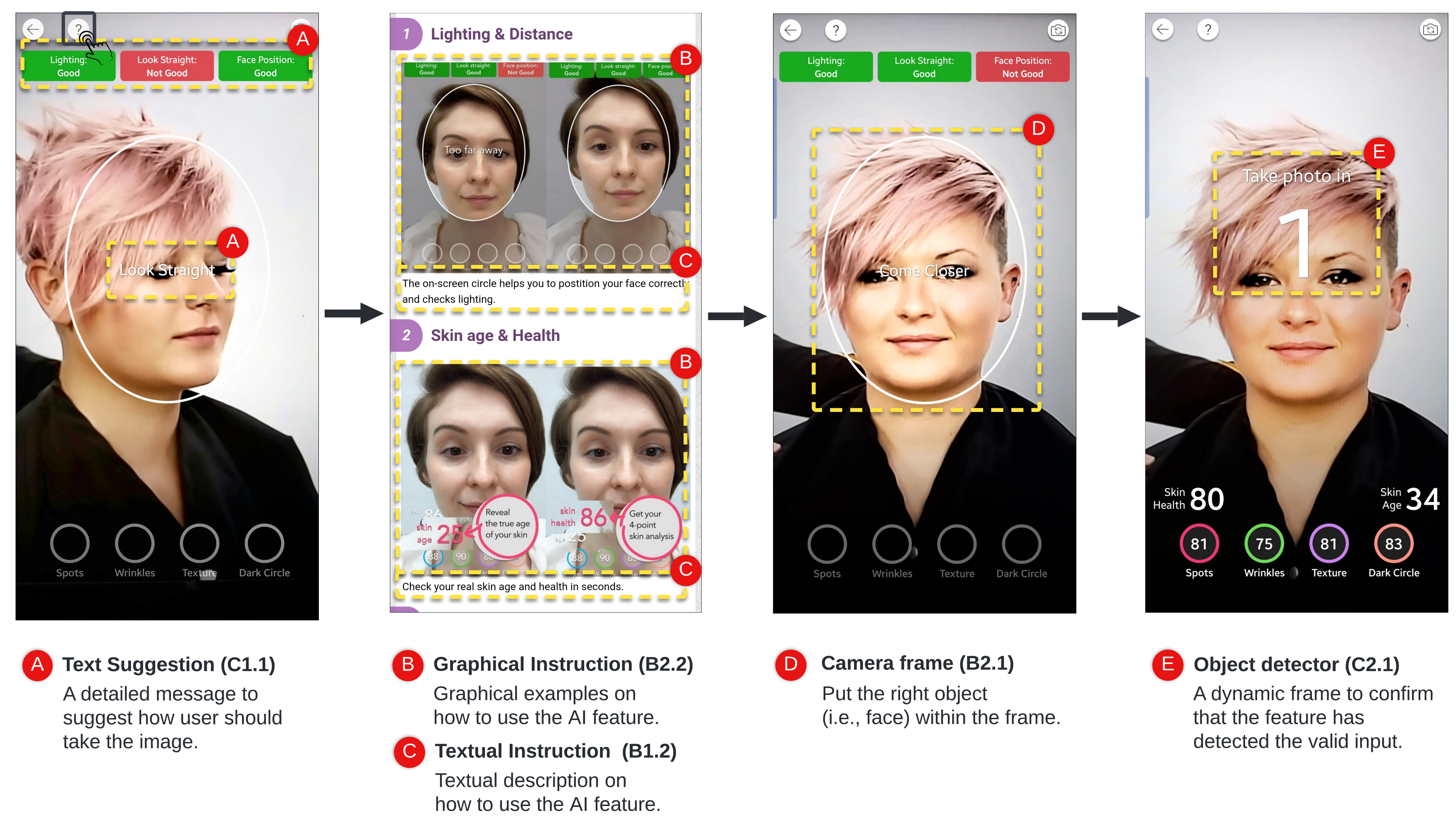}
	\caption{An example user-AI interaction on \texttt{YouCam Makeup}, one of the most popular makeover apps powered by AI (>100M) \cite{github_gallery} with a face detection model checking skin health and edits face makeup. From left to right, given an initial input, the app (A) first, specifies a certain aspect of the issues (e.g., positioning); second, advises users to read the illustrative instructions (B) and (C); third, while taking the input with a proper angle, suggests making the input with a closer distance (D); finally, a three-second countdown (E) confirms the valid input detected.}
	\label{fig:dynamic_int}
    \Description{An example of AI-human interaction patterns on YouCam Makeup, one of the most popular makeover apps supported with AI technology (>100M). A face detection model checks skin health and edits makeup on an image of the human face. From left to right, the interaction process begins: First, users provide a flawed input, so the feature used text suggestion patterns to specified a certain aspect of the issues (e.g., positioning); second, to correct the issues, the feature advised users reading the illustrative instructions with graphical and textual instructions; third, while taking the input with a proper angle, the feature suggested making the input with a closer distance using camera frame pattern; finally, a three-second countdown appeared indicated that the feature detected the valid input.}
\end{teaserfigure}

\maketitle

\section{INTRODUCTION}
\label{sec:intro}


Deep learning has shown its power in many domains, including object detection in images, natural-language understanding, speech recognition, etc~\cite{ren2015faster, devlin2018bert, gulati2020conformer}.
Meanwhile, smartphone apps (3.5 million in Google Play and 2.2 million in App Store~\cite{gPlaystoreAppNum}) have now become the most popular way of accessing the Internet as well as performing daily tasks, e.g., reading, shopping, banking and chatting.
To make machine learning more accessible to end-users and smartphone apps smarter, many deep learning models are embedded into smartphone apps for advanced tasks such as speech translation in Google Translate, grammar correction in Grammarly, image search in Naver (>100M) \footnote{See app link in our Github repository\cite{github_gallery}.}, etc. 
These AI-powered smartphone apps\footnote{We call ``AI apps'' for brevity within this paper} achieve fancy features to attract users, posting a competitive advantage over the markets.

Among these AI-powered smartphone apps, on-device AI apps have their unique advantages compared to the cloud AI apps.~\cite{dhar2021survey, xu2019first, sun2021mind,huang2021robustness}
First, on-device AI apps process data in local device, which could protect user data, while cloud AI apps rely on network calls and have to send user data to the cloud server, yielding many potential security issues like user privacy.
Second, on-device AI features remain usable while cloud AI features can not process data without the network.
Third, with the increase of computing capability of smartphones, on-device AI apps have quicker processing time than cloud-based apps.
Lastly, developers can also save budgets when deploying AI models in local devices because they do not need to rent a cloud server.
As the emerging way of deploying AI models into smartphones, the situation of AI features and their interaction designs on how to involve end-users remain unknown.


Since deep learning has already moved from being a back-end tool for enterprises to taking the frontline role within technology interfaces, there is a direct interaction between AI models and end users.
Different from conventional software written in rigid logic which specifies the clear input and corresponding output, AI or deep learning are relatively fuzzy i.e., reacting to changing information and conditions based on the data they receive (e.g., varied natural languages, different lengths of audio). 
Instead of designing specific reactions to a static set of scenarios, this dynamic behavior makes the UI/UX design of an AI app on a 6-inch-screen uniquely challenging.
In addition, as the performance of AI models highly depends on big data training, there is an output uncertainty if the model input is far different from the training data (e.g., too far from the camera and low-volume speaking). 
When designers proactively show users how to get good results (e.g., adjusting the lighting or moving closer to the camera), designers help them benefit from the feature and establish a more accurate mental model of the feature's capabilities~\cite{g_PAIR}. 
Such harmony between AI performance and user experience further aggravates the difficulty of interaction design of AI features.
Therefore, a great AI app depends on well-designed models as much as it depends on a well-designed UI and user experience.

As features powered by machine learning can be very diverse, design patterns may help make these experiences usable, beautiful, and understandable.
However, either most mobile app UI/UX design guidelines~\cite{material_io, ios_design} do not cover AI features, or general guidelines concerning human-AI interaction~\cite{microsoft, g_PAIR} do not specifically take mobile apps into consideration. 
Popular design-sharing sites (e.g., Dribbble) could provide conceptual inspiration, but it may be far from real-world implementation, and the lack of AI knowledge also creates a gap between designers and developers who develop the AI models.
These difficulties of UX design and the limitations of existing tools are confirmed in our informal interview with professional designers (Section~\ref{sec:motivation}).

To address these issues, we conducted a systematic analysis to understand the current status of AI features in the mobile market by investigating hundreds of real-world AI apps (Section~\ref{sec:RQ1}).
\jieshan{
In our work, we start with on-device AI apps because of two reasons.
First, on-device AI apps are gaining more attraction and become a promising direction for AI apps, and their interaction designs share the same rationale with apps with on-cloud AIs.
Second, on-device AI apps are much easier to be recognized than on-cloud AI apps because they include their AI models in the app package.
Since traditional methods can achieve similar functionalities as AI models, we need a reliable method to recognize whether an AI model is deployed or not.
For example, traditional image processing methods like SIFT~\cite{lowe1999object} can also be used to enable face ID and it is hard for us, as a user, to figure out the underlying algorithm by exploring its user interfaces. While some may consider analyzing the API calls to locate the potential usage of cloud AIs, it is in fact not feasible as the API name can be arbitrarily defined by the developers. 
}
We then observed how designers develop user-AI interaction in real-world AI-powered mobile apps and constructed a taxonomy of these design patterns(Section~\ref{sec:RQ2}).
As seen as a good example in Fig.~\ref{fig:dynamic_int}~\footnote{The example face is either from the app or a Youtube Video (https://www.youtube.com/watch?v=7wjFvfGYM74) which is licensed under Creative Common Attribution for free usage.}, there is a placeholder text to suggest input (A), and a detailed instruction page (B, C) pops up by clicking the help button.
When people interact with the feature, there is a camera frame (D) to remind users to put their faces into the area to achieve the best performance of the model.
The input is confirmed as valid with dynamic feedback without overwhelming them (E). 

As a result, we identified 255 AI features from 176 apps, comprising of 13 kinds of AI features (91.8\% are vision-related).
We then summarised \nIntPatternTotal{} examples into three primary types of interaction patterns, namely minimal feedback, general feedback, and fine-grained feedback, and ten secondary and nine tertiary patterns as seen in Fig.~\ref{fig:taxonomy}. 
We implemented a multi-faceted search enabled gallery \cite{gallery} for searching concrete examples of different kinds of AI models and interaction patterns.
Based on our findings and the gallery, we conducted a user study with six experienced UX designers to understand the usefulness and future applications of our taxonomy and gallery, which further proves the necessity and usefulness of our work.

As a summary, in this work, we make the following contributions:
\begin{itemize}
	
	\item To the best of our knowledge, this is the first work to explore fine-grained user-AI interaction patterns in mobile on-device AI apps. 
	
	\item We conducted a large-scale systematic analysis of \nApp{} AI apps from \nAppCollected{} real-world mobile apps on Android platform, which reveals the usage of on-device AI features and the practical interaction patterns for these AI features in the taxonomy and gallery\footnote{\url{https://algaespace.github.io/interaction_gallery/Gallery_static/index_page/index.html}}. 
	
	\item We carried out two rounds of interview with professional designers, which demonstrates both the limitations of existing tools, and the usefulness of our findings.
	
	
	
\end{itemize}

\section{RELATED WORK}
\label{ssect:Related_work}

Our work is related to two aspects,  i.e., UI/UX design in mobile apps and human-AI interaction.

\subsection{Conventional UI/UX Design for Mobile Apps}
\label{ssect:convent_design}
GUI provides a visual bridge between apps and users through which they can interact with each other. 
A good GUI design is difficult and time-consuming, even for professional GUI designers, as the designing process must follow many design rules and principles~\cite{handheld_mobile_device_guidelines}, such as fluent interactivity, universal usability, clear readability, aesthetic appearance, and consistent styles~\cite{prototype_fidelity, coursaris2011meta, miniukovich2015computation}.
To assist the mobile UI design, many studies are working on large-scale design mining including investigating the mobile UI design patterns~\cite{alharbi2015collect}, color evolution~\cite{jahanian2017colors, jahanian2017mining}, and UI-related users' review~\cite{fu2013people, martin2017survey}.
As dominant OS vendors of mobile apps, Google establishes Material Design~\cite{material_io} and Apple develops Human Interface Guidelines~\cite{ios_design}, which have also implemented comprehensive guidelines to support the app interface design.

However, most of these works do not cover AI elements, therefore the findings cannot be directly applied to AI apps.
The conventional UI/UX guidelines were built based upon fixed input, output and machine behaviors~\cite{transitioning}, whereas AI-based apps accept the users' unpredictable behavior and return fuzzy results. 
In contrast, our study aims to develop an empirical study specifically of interaction design patterns for AI features in mobile apps.


\subsection{Human-AI Interaction}
\label{ssect:ai_human_interaction}
As AI technology becomes more ubiquitous, many researchers have noticed challenges in the system design and ML development including a lack of dialogue f4rdc\cite{role_yang} and a skill gap between UX and ML expertise \cite{liao2020questioning}.
UX designers are encouraged to increase their literacy in the area of machine learning, and work with it as a design material~\cite{ux_design_as_ML_mat}.
For example, they should have an understanding of what AI can and cannot do to a system, and the role of feature engineering in ML models~\cite{investigating_yang, Profile_AI_Yang, IBM_question_driven, ux_design_as_ML_mat, holmquist2017intelligence, yang2018mapping}.
Researchers \cite{re_examining} elaborated that capabilities and limitations can mean what AI can afford (e.g., recommendation features), how well it performs, and the kinds of errors it produces. 
They mentioned that the early design ideation stage has the most uncertainty because designers try to understand what design possibilities AI can generally offer without a catalog of available AI capabilities.
As such, the interaction design can take the back-end model's performance into account\cite{investigating_yang, role_yang}. 

In addition to individual research studies, general guidelines concerning human-AI interaction are also established in both academics and industry including, Berkel's categorisation of Human-AI Interaction \cite{berkel}, Shneiderman's two-dimensional framework of Human Centered Artificial Intelligence (HCAI)\cite{shneiderman2020human},   Microsoft's 18 design guidelines for human-AI interaction\cite{microsoft}, Google's People + AI Guidebook~\cite{g_PAIR} for designing with AI.
\jieshan{For example, Microsoft's design guidelines provide general guidelines in terms of all AI features but they do not consider the differences among different AI features and can not provide detailed, operationalizable and concrete guidance on what to design and how to engage users in a real-world scenario.}
Specifically for mobile app development, conventional Google material design adds simple guidelines for designing interface for AI apps such as object detection and barcode scanning \cite{M_IO_ML}.
\jieshan{While these guidelines are useful, it is hard for designers to understand other AI features and extend the underlying patterns to their designs as Google only provides two examples.}
Apple also introduced basic guidelines to help design the UI/UX of a machine learning app in areas such as getting feedback, displaying data, handling mistakes, and enabling corrections~\cite{apple_ML_gl}.
\jieshan{While they provide some high-level instructions, how to implement and transfer these guidance into practical and real interaction design remains unknown and unsure.}
Within the community, some AI feature interaction designs can also be seen in design-sharing sites like Dribbble. 
\jieshan{However, while they provide many examples, for designers who are not familiar with AI products, they may find it hard to understand the underlying rationale and thus hard to apply and generalise to their own design.}

Moreover, much space is not covered by the aforementioned research studies and guidelines.
First, the deployment of AI models into mobile apps and corresponding interaction design are rarely explored in these works.
Second, most guidelines propose general rules for human-AI interaction, and they cannot be directly applied to the context of AI human interaction in mobile apps.
Towards those limitations, we are specifically targeting at a systematic analysis of user-AI interaction in 6-inch-screen smartphones by investigating hundreds of practical AI apps from the largest app store in the world.
Different from prior work on abstract or concept design guidelines, all findings in our study are from real-world practices with detailed examples which are easy to understand for designers and developers who are even familiar with AI technologies. 
\jieshan{We summarised an interaction pattern taxonomy to distill the underlying rationale from these existing interaction designs to guide and help designers.}

\section{MOTIVATIONAL STUDY}
\label{sec:motivation}
In this section, we conducted interviews with several senior UI/UX industrial practitioners to understand their real need and challenges to motivate our study.

\subsection{Interview setup}
To get feedback from designers about their needs for developing AI software, we conducted an informal interview with four professional designers responsible for design innovation, including two product designers from Microsoft, Atlassian, and two UI/UX designers from PwC and Huawei.
All participants have been working on software design for more than 5-year with modest working experience on AI-related products.
Each interview was conducted over a video call and lasted about half an hour.
We first showed them screenshots of AI features from at least five apps as the warm-up.
Then we asked them about their working experience in designing AI product, and the challenges during that process, especially when they were designing mobile apps.

\subsection{Findings}
\textit{Feature familiarity for end-user.}
Compared with conventional software, the learning curve of using AI models is much steeper, especially in judging the validity of the input.
We also found that AI technology could bring challenges to end users. One designer said \textit{``Another challenge is the acceptance of new technology for users.  Because the users get used to using the text inputs (i.e., traditional interaction), so it takes longer to ascertain the scenarios and usage of voice inputs with the AI feature.''} She further elaborated \textit{``So the solution is to mix the traditional interaction method and AI-driven method together (e.g., putting text input, image input \& voice input together). This boosts user confidence because the user can choose their preferred way of input.''} Interestingly, another designer thought that AI technology could bring challenges to other targeted users who have specific requirements (e.g., disabled group). Therefore, in spite of the opportunities for AI technology, there are drawbacks for end users in using this unfamiliar interaction.

\textit{Insufficient knowledge to have a proper expectation of AI features.}
Participants from our user study further validated the knowledge gaps \cite{ux_design_as_ML_mat, investigating_yang, yang2018mapping} between AI engineers and UX designers.
One participant noted that \textit{``It is hard for designers like him from non-technical backgrounds to understand AI expectations and what a regular interface can do to convey that expectation.''}
Another designer further elaborated that\textit{ ``The novelty of AI products made us hard to understand the product requirements. For example, without talking to AI engineers, I would not understand that the performance of AI features heavily relied on the user input and behaviours.'' }
Therefore, if designers did not understand the expectation of AI models in back-end, designers might not take model training into a consideration for their interaction design. 

\textit{Insufficient tool or gallery for designing interaction of AI features.} When reflecting on current practices on UI/UX design, participants often noted that they would use design sharing platforms such as Dribbble, Muzli 
, and Mobbin to get inspiration. One explained that \textit{``These examples are conceptual which is hard to gauge the practicality of design idea.''} To realise the feasibility, one mentioned that they had to ask developers with ML backgrounds. This brought communication issues because the focus for designers and developers was on the visual designs and model performance respectively. Another designer explained that \textit{``We have some internal tools for supporting the design of AI; however, we had to manually find similar apps, download and explore them individually which were time-consuming and labor-intensive.''} All participants emphasised the importance to get inspiration by sharing design knowledge, creativity, and aesthetics via these platforms. However, from this study, we learnt that there is a lack of tools for the actual implementation of real-world apps' interaction designs.


Based on the challenges we identified in the formative interviews, we found the design goals of the empirical study i.e., outlining the current status of AI usage in real-world mobile apps, and the corresponding design patterns used to interact between end users and AI models.
With such observations well accumulated, organized and displayed, a gallery may help inspire designers to design the AI-user interaction in their working apps.  
\section{METHODOLOGY FOR EMPIRICAL STUDY}
\label{sec:methodology}
\begin{figure*}
\center
    \includegraphics[width=1\linewidth]{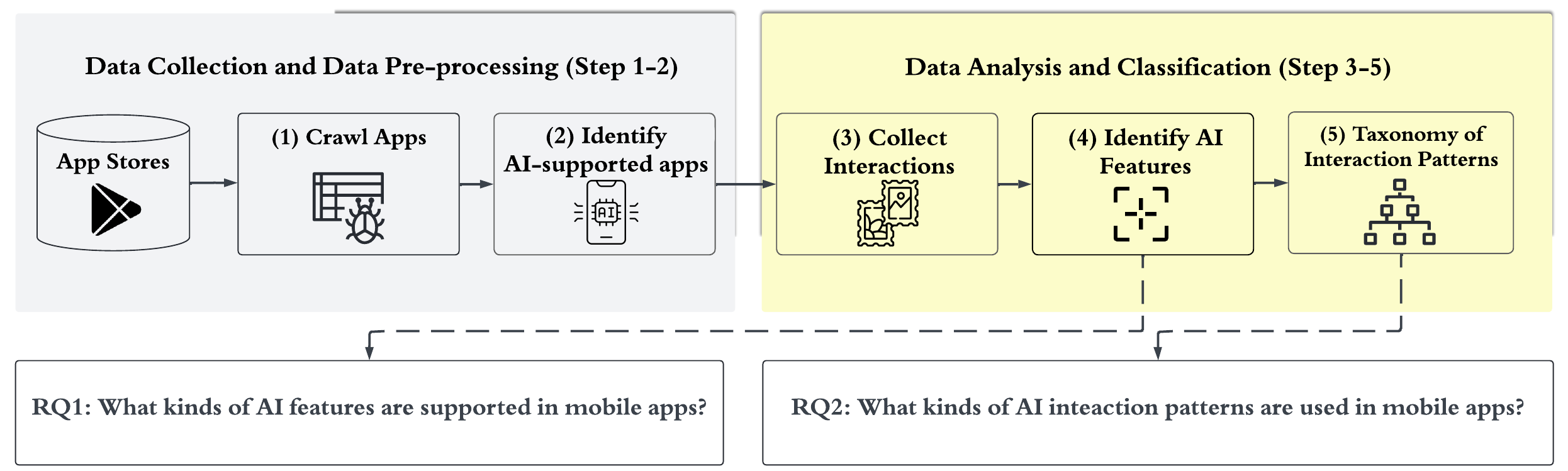}
    \caption{Overview of the methodology for empirical study. }
    \Description{Our methodology for the empirical study consists of three main blocks. The first block is named as ``Data Collection and Data Pre-processing (Step1-2). We crawled mobile apps from the app store and then identified the AI-supported apps from them. The second block is called ``Data Analysis and Classification (Step3-5). After we got the list of AI-supported apps from Block one, we collected interactions from these apps. Step 4 and 5 are to identify AI features and summarise a taxonomy of interaction patterns to answer our two research questions. Block 3 shows our research questions: RQ1 is "What are kinds of AI features supported in mobile apps?"; RQ2 is "What are kinds of AI interaction patterns used in mobile apps?" }
\label{fig:overview_me}
\end{figure*}

Based on the design goals, we carried out a large-scale systematic analysis of mobile apps with AI-based features to answer two research questions (RQs) i.e., \textit{What kinds of AI features are there in mobile apps?} (RQ1-Section \ref{sec:RQ1}), \textit{what interaction patterns are used in mobile AI apps?} (RQ2-Section \ref{sec:RQ2}).
The overall approach is illustrated in Fig.~\ref{fig:overview_me}.
In detail, we first crawled a large-scale dataset of mobile app packages, identified AI-supported apps and filtered out unusable apps.
Two authors then installed apps, used the AI features in these apps, and recorded the interactions in these AI features.
Based on the collected interactions, we open-coded the taxonomy of the interaction design patterns iteratively and cultivated our terminology as shown in Fig.~\ref{fig:taxonomy}.



\subsection{Data Collection}

\subsubsection{Step 1 -- collecting an initial set of Android mobile apps}
To get large-scale mobile apps, we focus on Android apps for two reasons.
Firstly, according to Statista Research Department~\cite{androidMarketShare}, Android operating system occupied 87\% share of the global market in smartphones in 2019, and is expected to increase over time. 
Secondly, the meta information and app packages (APKs) of Android apps can be easily obtained via Google Play store and some third-party websites, which enables us to collect a large-scale dataset and perform analysis.
Hence, for the preparation of the dataset used in this work, 
\yujin{we first select 62,822 apps across 49 categories\footnote{The covered app categories may be different from the ones on Google Play due to updates.}, such as entertainment, education and business, from Google Play.
These apps include AI apps identified in existing AI app-related studies~\cite{xu2019first,sun2021mind,deng2022understanding,huang2022smart} and the trending apps of each category.
We then crawl the metadata (e.g., app name and description) of the selected apps from Google Play and utilize them as seeds (i.e., inputs to our Android app crawler) for collecting corresponding APKs through third-party websites including APKMirror~\footnote{https://www.apkmirror.com/}, APKPure~\footnote{https://m.apkpure.com/} and APKMonk~\footnote{https://www.apkmonk.com/}.}

\subsubsection{Step 2 -- identifying AI-supported apps}
The second step is to identify the deep learning models from the collected APKs, and remove duplicate or unusable apps.
As we have collected a large-scale dataset, it is infeasible and time-consuming to manually go through each app to check if they contain AI features or not. Such a manual method cannot be scaled up. Therefore, we proposed to use an automated method to detect AI apps by checking the model files in app packages.

Inspired by the popularity of on-device AI models~\cite{xu2019first} (i.e., deploying deep learning models into the local app), we decompiled the app and searched for on-device AI model files by filename extension \footnote{see supplementary materials}.
If any such exists, we tagged it as an AI-supported app for subsequent use.
\jieshan{Note that while some AI-powered apps rely on network calls to perform inference, it is infeasible to recognise these network calls by analysing their decompiled source code because there is no standard of how to define the API name which triggers the cloud AIs. Therefore, developers can define any API names they like, and we could not establish some name patterns to locate the potential usage of AI models.}
During identification, we used the package name of each AI-supported app as the unique identifier and recorded it to avoid duplication (i.e., an AI-supported app is tagged only if its package name is not in our records) because our crawler may download certain apps more than once due to network anomaly.
After this step, we obtained 335 AI-supported apps.

To ensure the correctness of identified apps, we manually installed the remaining apps for subsequent AI feature identification.
By inspection, we noticed that 159 apps are unusable for two reasons: 
(1) 5 app providers terminate the services; 
(2) 154 apps cater to specific audiences like corporate employees or owners of external Internet of Things devices (e.g., smartwatch), all of which require login credentials for further use. 
Thus, such apps are removed to facilitate the collection of human-AI interactions.
We got 176 AI-supported apps as our target.
We provide the details of these app names and metadata in supplementary materials and the GitHub repository~\cite{github_gallery}.



\subsection{Data Analysis and Classification Method}
In this section, we first collected interactions in each app.
Based on the collected recordings, we identified the initial set of AI features by the filename of AI features, and then adopted open coding approach to further refine the type, and usage of AI features to answer RQ1, and cultivate our taxonomy for RQ2.

\subsubsection{Step 3 -- collecting interactions.}
After the previous two steps, we now obtained our final set of AI-powered apps.
To understand the interaction designs and curate a taxonomy for future use, we then ran each app and recorded the interactions.

To do this, we split the final set of apps into two splits, and two authors interacted and recorded with the AI features of one split.
Each author first recorded interactions in three randomly sampled apps, and discussed the collected recordings together to cultivate a standard for the recordings.
After discussion, these authors agreed on the following standards. 
First, to ensure the comprehensiveness of the collected interactions, for each app, the authors should first read the app description in Google Play, and install and use the app manually for its intended usage for at least 10 minutes.
Second, after the authors are familiar with the app features, they then record the interactions.
Each app will yield one recording.
A recording should start with the triggering page of the AI features, i.e., the page that triggers that AI feature, and end with the result of the AI feature. 
As one AI feature may enable different input methods, all input way should be recorded in one video. For example, as an image input, the authors can either choose a photo in their album or take a new photo via camera. 
For each AI feature, the authors are asked to try several times by using valid and invalid inputs to see the interactions. 
All interactions are collected using Google Pixel 5, which has a dual-lens camera, 8 GB of memory and a CPU of Qualcomm Snapdragon 765G SM7250-AB (7nm). 

\subsubsection{Step 4 -- identifying initial set of AI features.}
To understand the kinds of existing AI features in mobile apps, we identified the AI features by using the hints from their embedded model filenames and by going through all recordings.
We found that the model filenames of the AI features follow some patterns that indicate their potential usage.
For example, one face detection model file is named as ``Contours.tflite''.
Therefore, we used these patterns to create the initial set of AI feature types to ease the AI feature identification process.
We then refined and added/removed the types of AI features by going through all recordings in the next step.


\subsubsection{Step 5 -- forming a taxonomy from existing Human-AI interaction designs}
To recognize and summarize the reusable patterns from existing human-AI interaction designs, we adopted an iterative open-coding approach. 
We made three steps through our collected interaction recordings.

The goal of the first step is to create the initial interaction coding set.
In this step, we first referenced existing Human-AI interaction guidelines, such as Google Material.io \cite{M_IO_ML}, Google's PAIR~\cite{g_PAIR}, and Microsoft's Human-AI Interaction Guidelines~\cite{microsoft} to identify potential analytic directions. 
We then went through recordings of 20 randomly sampled apps. 
Based on the initial set of AI feature types from Step 4 and the insights drawn from existing guidelines, two authors independently coded the potential dimension and properties by watching each interaction recording of each app, and noted any terms that are not part of the initial vocabulary.
After the initial coding, the researchers discussed the discrepancies and the set of new terms until a consensus was reached.
This step yielded three primary dimensions, including types and usage of AI features (e.g., detecting the existence of faces), interaction categories (e.g., minimal feedback), interaction styles (e.g., on-screen text feedback and vibration), and some sub-types of each dimension. 

In the second step, two authors checked the rest interactions individually and identified the reusable characteristics of each dimension based on the initial taxonomy. 
If there are new dimensions or more sub-types identified during this process, the authors will note them.
After scanning and identifying the properties, these two authors first compared the updated taxonomies, discussed the differences and used the interactions as examples to find an agreement on the taxonomy. After obtaining an agreement, we formed our final taxonomy. 
During this process, we tried to make sure the terms used are extensible and explanatory so that future studies can reuse our taxonomy.

Lastly, these two authors went through all interactions together and refined the initial labeling.
If they can not make an agreement, a third author will involve.
The final decision on the classification was taken by majority voting. This was done together to mitigate the risk of potential bias.
Note that after those processes, we also further refined the terms based on the feedback from the industrial practitioner during the user study as we will describe later (Section~\ref{sec:RQ3}).
Additionally, we also counted the number of each pattern in each AI feature after removing duplicates. 



\section{RESEARCH QUESTIONS AND FINDINGS}
\label{sec:findings}

In this section, we reported our findings to answer two research questions proposed in Section~\ref{sec:intro}.
First, we examined the functionalities and usage of AI feature in the mobile apps we crawled to answer RQ1 in Section \ref{sec:RQ1}. 
We then investigated how AI features interact with users, and summarised the patterns into a fine-grained taxonomy comprising 22 categories of different levels to answer RQ2 in Section~\ref{sec:RQ2}. 


\begin{figure}
    \begin{subfigure}[ct]{0.7\linewidth}
        \includegraphics[width=\linewidth]{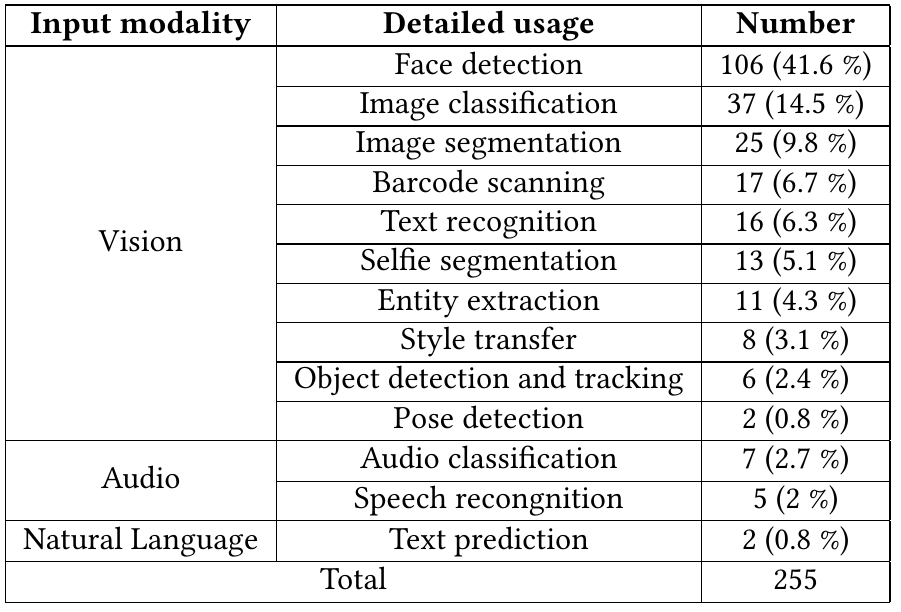}
        \caption{The number of AI feature usage.}
        \label{tab:model_clas}
    \end{subfigure}
    \hfill
    \begin{subfigure}[ct]{0.98\linewidth}
     \includegraphics[width=\linewidth]{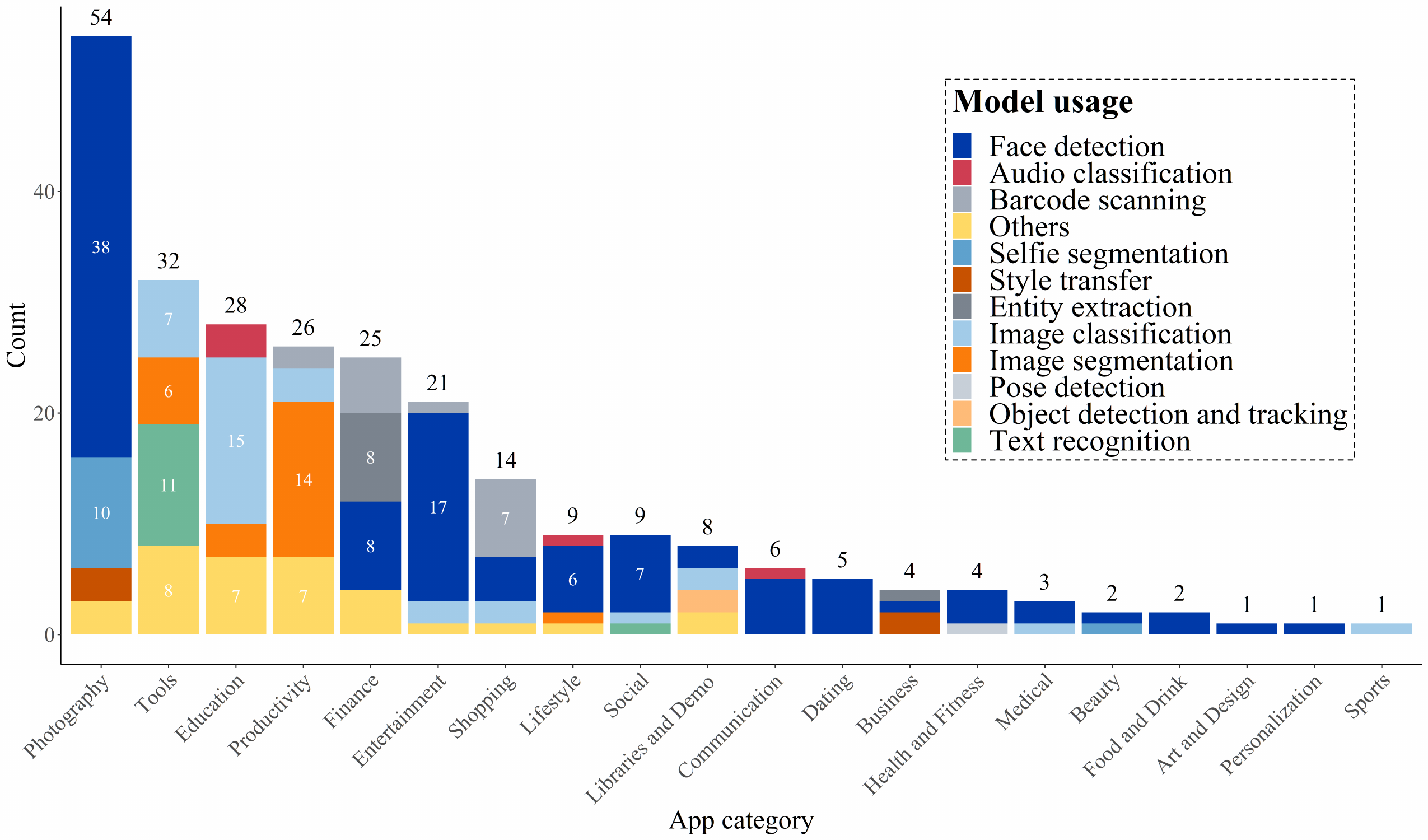}
        \caption{The number of AI features in different app categories.}
        \label{fig: model}
    \end{subfigure}%
    \vspace{-3mm}
    \caption{Statistics of the types and usages of AI features.}
    \Description{This figure shows the statistics of the types and usages of AI features. It contains two subfigures from left to right. The first figure is a table showing the number of detailed AI feature. It has three columns, namely input modality, detailed usage and number. The second figure is a stacked barchart of the number of AI feature in app categories containing AI-supported apps. Among all app categories, Photography app category contains the most AI features (54) in our dataset, almost the twice as the number of the second most app category (Tool). As for the usage of AI features, 16/24 app categories have the face detection features dominated. Other app categories like Tool, more focus on the organizing documents so it uses text recognition and image classification more often.}
    \label{fig:RQ1}
\end{figure}

\subsection{RQ1: What Kinds of AI Features are Supported in Mobile Apps?}
\label{sec:RQ1}




Overall, we identified 255 AI features in \nApp{} apps, comprising of 13 unique AI features, and each app contains an average of 1.4 AI features.
These apps spanned 20 app categories, ranging from Finance to Photography, and supported various AI features, such as beautification using face detection features, and auto-crop documents using image segmentation features. 
Such diverse app categories and functionalities indicate that AI techniques are promising techniques and attracting more companies to utilize them to enrich their app functionalities.
For those categories without AI features, the reasons may be that functional categories like Parenting are commonly used for information sharing and tracking purpose, and entertaining categories like Arcade are primarily based on in-app presets. 

In detail, these AI features can be broadly categorised into three types in terms of their input modality, namely, vision-based, audio-based and natural-language-based AI features.
As seen in Table~\ref{tab:model_clas}, among these three types, 91.8\% AI features are vision-based.
This is reasonable as computer vision techniques are now mature, with many ready-to-use pretrained and lightweight models of high accuracy and can inference in a short time\cite{iandola2016squeezenet, howard2017mobilenets, lecun2015deep}.
For example, 54\% of 39 problem domains published in Tensorflow hub \cite{tfHub_domain} are vision-based (i.e., 17 being image-related and 4 being video-related). 
Moreover, many platforms also provided some ready-to-deploy machine learning model APIs for the app developers.
For example, 8/12 APIs in ML-Kit \cite{ml-kit} are vision APIs.
\jieshan{In addition, research on CV model compression is also more mature than works on other types of models, which ensures acceptable app package sizes and efficient inference, and it leads to more adoptions of CV models in mobile apps. }
Specifically, face detection features account for the highest proportion of vision-based AI features. 
It is commonly found in the Photography app category, in particular camera and photo beauty apps, where users’ faces are auto-detected and then beautified based on user requirements. 
In comparison, pose detection features are the least used techniques (\fpeval{round((2 / \nIntPatternTotal{}),3) * 100}\%) because of its complexity, and these AI features are normally used to evaluate and correct posture.


In addition, 12/255 (4.7\%)  AI features are audio-based, which are either used to classify the audio or recognise the speech content to aid speech-to-text translation.
For example, an app named \texttt{Sleep as Android: Smart alarm} \cite{github_gallery} specifically informs if users talk and snore while sleeping, and \texttt{Naver} \cite{github_gallery} enables voice search given the user voice input.
Lastly, we only found 2/255 (0.8\%) natural-language-based AI features in our dataset, which are used to suggest some texts when users are typing.



We further analysed the relationship between app categories and AI features. 
As seen in Fig~\ref{fig: model}, AI features exist widely across different app categories. 
In our dataset, the top app category using AI features is \textit{Photography}, with a number of 54 AI features (\fpeval{round((54 / 255),3) * 100}\%), nearly twice as the number in the second most app category (\textit{Tool}). 
This app category is mostly about auto-locating face contour in facial beauty apps, in which the facial recognition technology is rather mature and hence widely used. 
In comparison, eight app categories, including \textit{Dating}, \textit{Business}, \textit{Health and Fitness}, \textit{Medical}, \textit{Beauty}, \textit{Food and Drink}, \textit{Art and Design}, \textit{Personalization} and \textit{Sports}, only contain less than or equal to five apps that deploying AI features.
The reason accounting for that phenomenon may be that the usage of AI technology is not yet aware in these domains. 
In terms of the types of used AI features, the face detection model dominates the usage in most app categories.
For app categories \textit{Dating}, \textit{Food and Drink}, \textit{Art and Design}, and \textit{Personalization}, we only recognize the usage of face detection features.
In contrast, only 4/20 app categories do not adopt such features.
For example, apps in \textit{Tools} category are mainly about utilities like translations of documents and texts, or storing and organizing documents and photos, so it uses text recognition and image classifications more frequently. 
All these findings show great potential for developing new apps with AI features in the future, and thus the quality of interaction design between end-users and AI features will become more important.

\subsection{RQ2: What Interaction Patterns are Used in Mobile AI Apps?}
\label{sec:RQ2}
\begin{figure*}[htbp!]
  \centering
  \includegraphics[width=1\textwidth,scale=0.99]{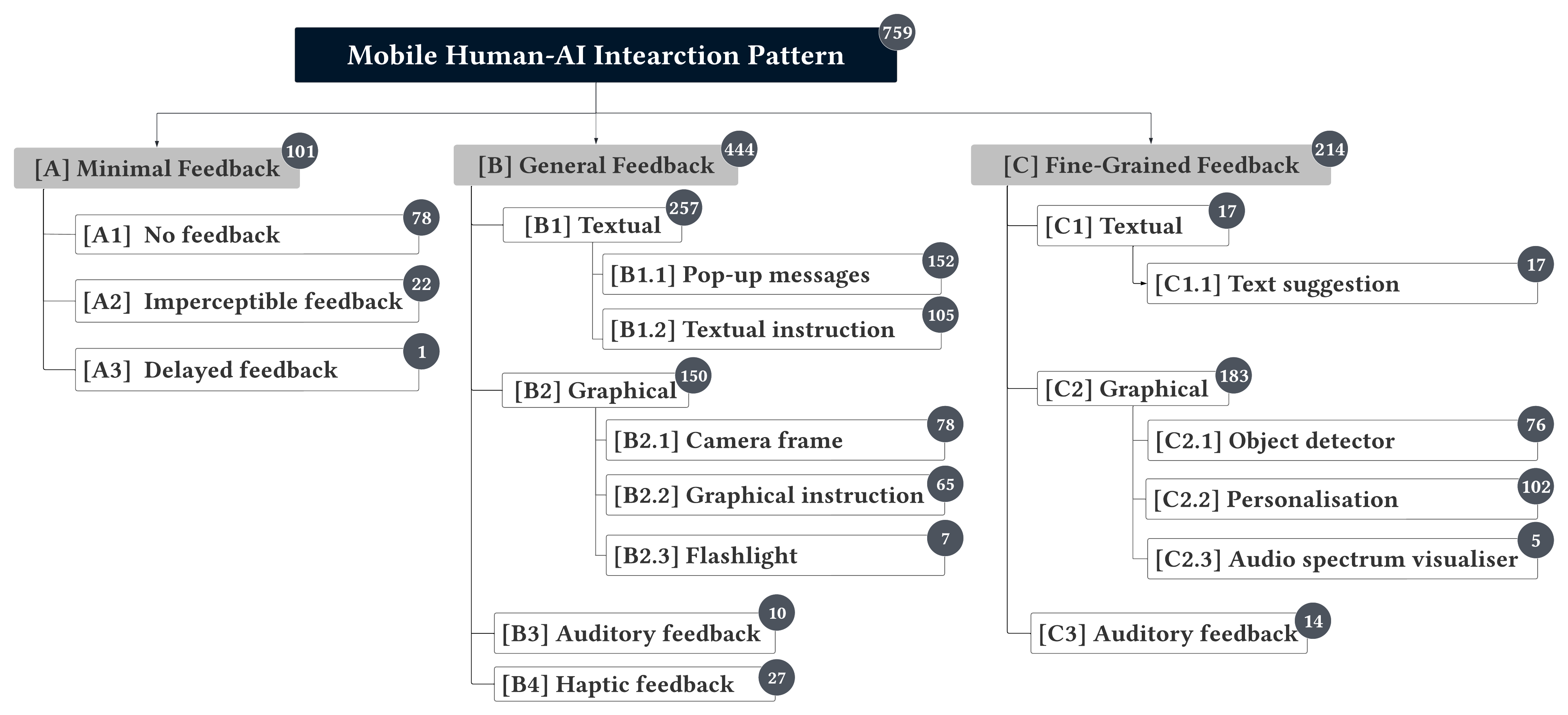}
  \caption{Taxonomy of user interaction with AI features. The number on the top right corner of each cell refers to the number of that pattern in our dataset. Note that one AI feature may adopt multiple interaction patterns. 
}
  \Description{The taxonomy is shown in a tree structure. The top layer is "Mobile Human-AI Interaction Patterns", followed by three primary patterns, namely [A]Minimal Feedback, [B]General Feedback and [C] Fine-grained feedback. Type A minimal feedback consists of three secondary patterns; Type B general feedback contains 4 secondary patterns and five tertiary patterns; Type C fine-grained feedback contains three secondary patterns and four tertiary patterns.}
  \label{fig:taxonomy}
\end{figure*}

Through our recursive analysis of our collected interactions from existing apps, we identified three primary interaction patterns that are used to present feedback to the end-users: minimal feedback, general feedback and fine-grained feedback.
For these primary patterns, we also identified in total ten secondary and 
nine tertiary patterns as seen in Fig.~\ref{fig:taxonomy}.

\subsubsection{\textbf{Minimal feedback (A)}}
\label{minimal_feedback}
is defined as imperceptible, trivial or delayed feedback the app responds to user actions. 

\underline{No feedback (A1)} means that there is no instruction or feedback given to users. 
The user can only recognise potential issues by themselves when they get unexpected final results from the AI feature, and then try to locate the problem and fix it (see Fig~\ref{fig:A.1}).

\underline{Imperceptible feedback (A2)} directly provides modified results processed by the AI feature to the users, and the users may not recognise or perceptualise such AI behaviours.
For example, Weibo~\cite{github_gallery}, a social media app, provides an AI feature that automatically detects any illegal or abnormal activities on the platform (e.g., abusive language), and then removes these comments or blocks the poster directly without user consent or confirmation.
While this feedback supports online safety, unreliable AI features may make bad decisions by hiding some benign comments.

\underline{Delayed feedback (A3)} intends not to interrupt users, but allows users to backtrack the results later.
A good example is the \texttt{Smart alarm} \cite{github_gallery} app, which automatically records and detects sleeping sounds (e.g., sleep talking and snoring). 
When it detects some sleep talks, it will quietly save results without notifying users instantly, and users can check the summary after they wake up. 

\textbf{Statistics:}
As observed, minimal feedback accounts for 13.31\% (101/759) of total identified patterns, and 97/255 (39\%) AI features adopt such patterns.
Among such feedback, no feedback patterns (A1) contribute the most (78/101-80.4\%), while imperceptible feedback (22/101-22.7\%) and delayed feedback (1/101) are less adopted.
The reason for this may be that certain apps such as selfie segmentation and face detection directly provide outputs to end-users without any specific input requirements.

\subsubsection{\textbf{General feedback (B)}} is defined as a general form of communication. Fig~\ref{fig:taxonomy} illustrated that it takes mainly four forms (i.e., textual, graphical, aural and haptic) to users. 

\begin{figure*}
\centering
\subfloat[No feedback (A.1).]{\label{fig:A.1} \includegraphics[width=0.20\textwidth]{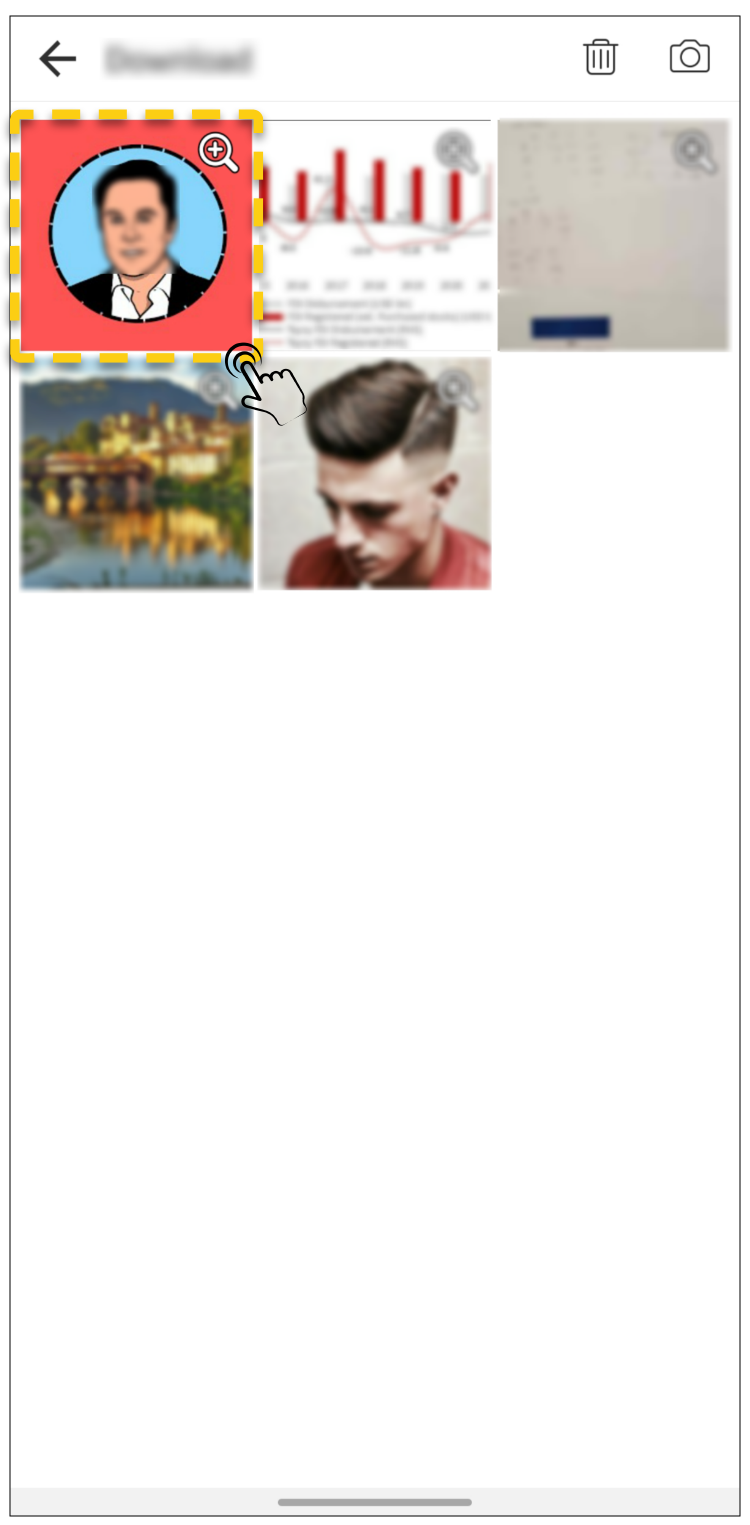}}
\hfill
\subfloat[A pop-up message (B1.1).]{\label{fig:B1.1} \includegraphics[width=0.20\textwidth]{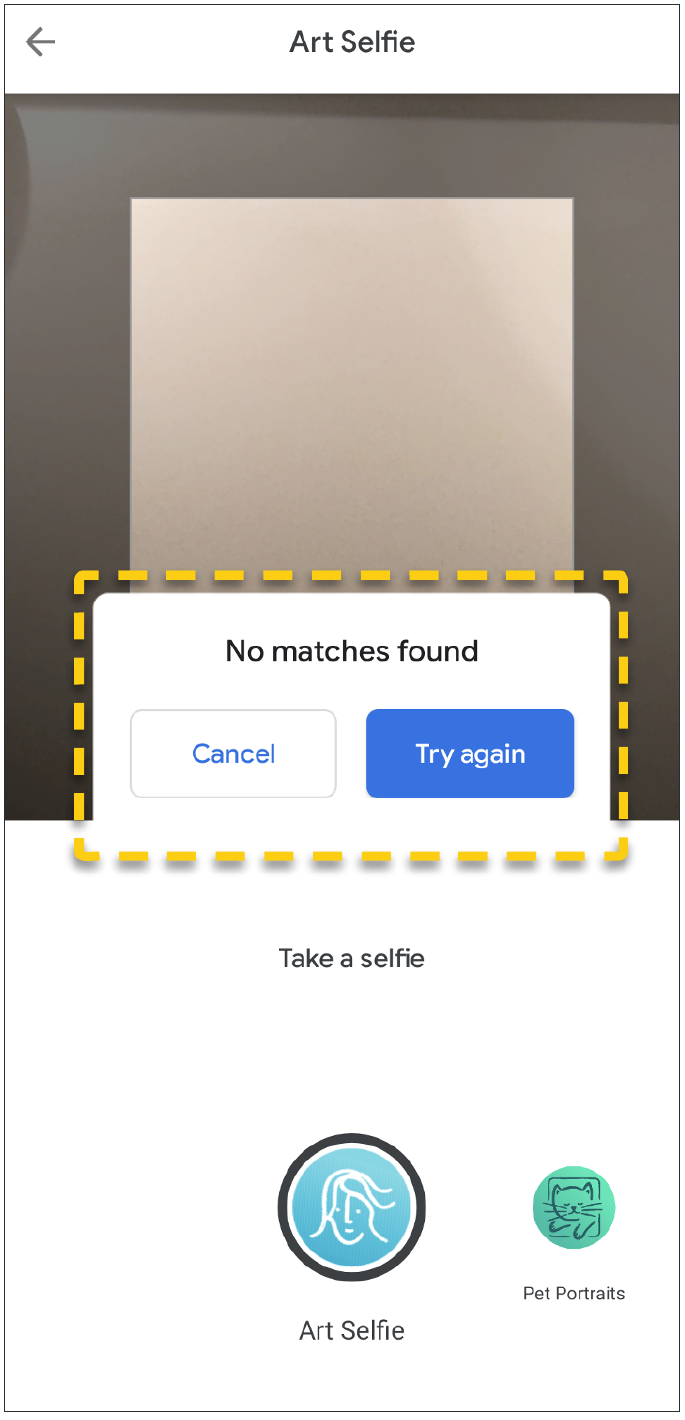}}
\hfill
\subfloat[An instruction with both textual (B1.2) and graphical (B2.2) elements.]{\label{fig:B1.2} \includegraphics[width=0.20\textwidth]{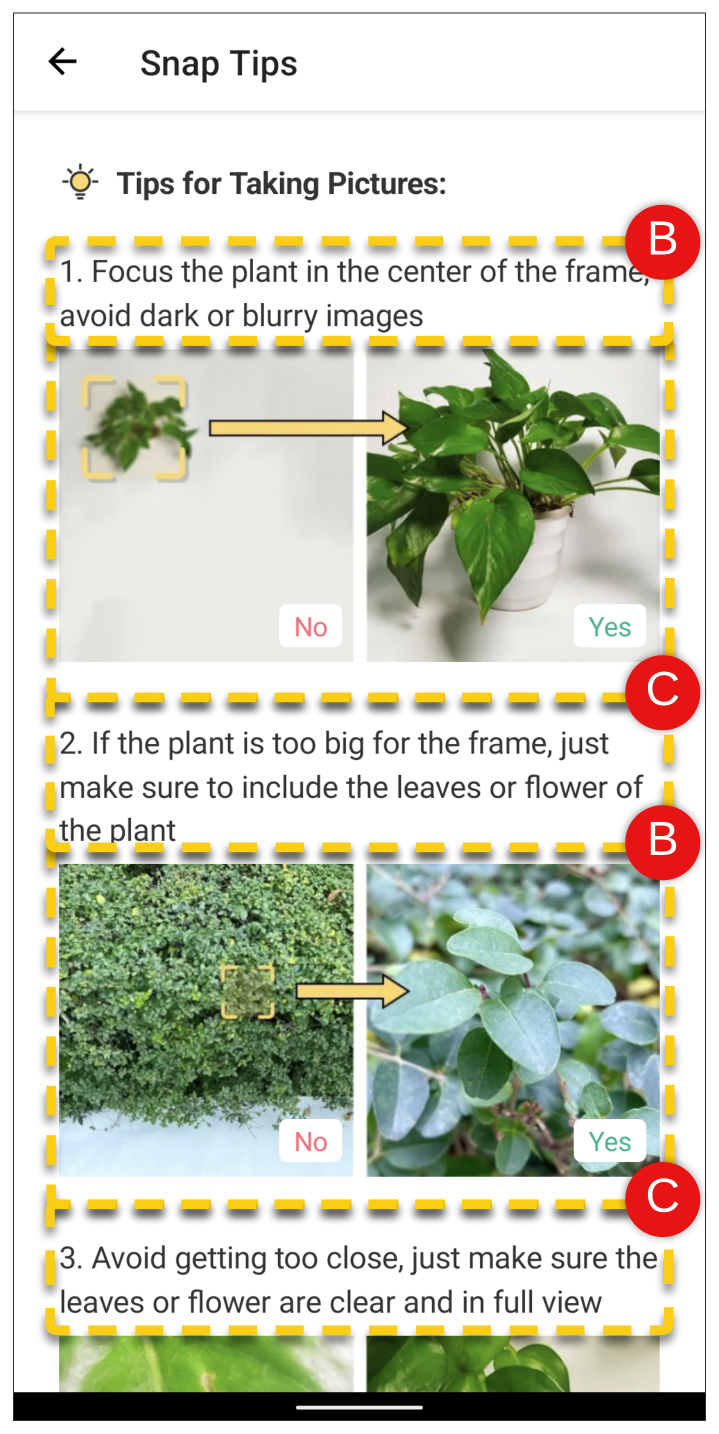}}%
\hfill
\subfloat[A camera frame and mushroom silhouette (B2.1).]{\label{fig:B2.1} \includegraphics[width=0.20 \textwidth]{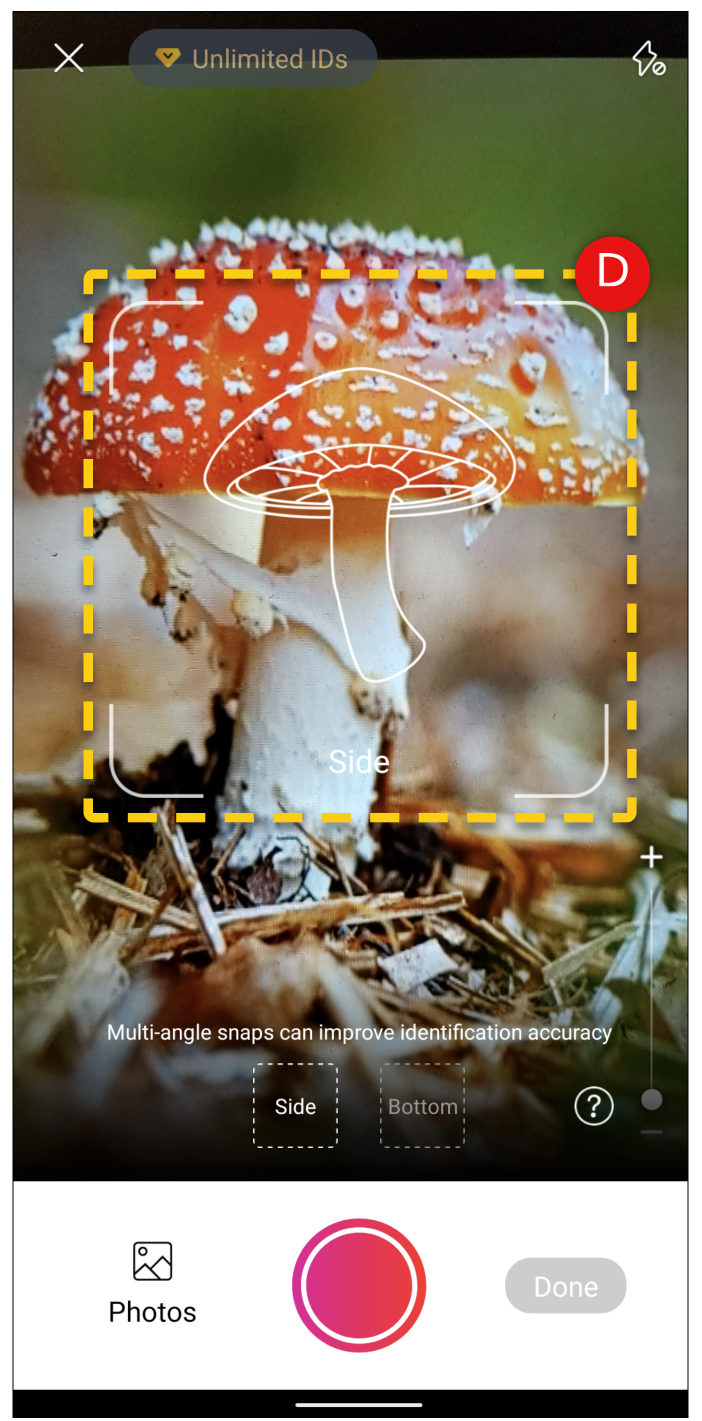}}%

\caption{
Example of interaction pattern A and B. Fig (C) depicts that instructions with some graphical elements shown in \raisebox{-0.5ex}{\includegraphics[width=3ex ,height=3 ex]{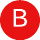}} on guiding how users should make correct inputs (i.e., plants), along with textual instruction shown in \raisebox{-0.5ex}{\includegraphics[width=3ex ,height=3 ex]{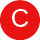}}. In Fig (D), a camera frame shown in \raisebox{-0.5ex}{\includegraphics[width=3ex ,height=3 ex]{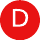}} hints users that they should take the input with a silhouette mirroring a mushroom within the frame.
}
\Description{This figure shows some examples of interaction patterns. From left to right, Figure(a) shows an example of no feedback pattern (A1) where users are asked to upload an image to the app. No hints or other feedback from the app. 
The second example shows an example of pop-up message (B1.1). The UI contains a pop-up message saying "No matches found" with two options "cancel" and "Try again. The third example shows an example of textual and graphical instructions given by a plant detection app. The last example include a mushroom silhouette that hints users to place the mushroom to the this area. }
\label{4figs1}
\end{figure*}

\underline{Textual (B1)} patterns use text elements to convey some guidance to help users understand the usage of AI features or give error messages, such as some \textit{pop-up messages (B1.1)} or \textit{textual instruction (B1.2)}.
One such example from \texttt{Google Arts \& Culture}\cite{github_gallery} \textit{pops up} an error message to inform users to have another try when the AI feature fails to detect the matches (Fig~\ref{fig:B1.1}).
Another example is that the app in Fig~\ref{fig:B1.2} gives \textit{textual instructions} to educate users on how to use the provided AI feature. 
It is usually accompanied by the graphical patterns (B2.2) as we will discuss later.

\underline{Graphical (B2)} patterns instead leverage graphical elements to provide illustrative examples to guide users.
Some add a \textit{camera frame (B2.1)} to enable end-users to put the right object in the right position when using camera. 
For example, Fig.~\ref{fig:B2.1} shows a rectangle frame 
with a mushroom silhouette to indicate that the target object is a mushroom, and the user should position the object inside the frame.
Some utilize illustrative icons and pictures, along with textual elements, to give better intuitive \textit{graphical instructions (B2.2)} to end-users (Fig.~\ref{fig:B1.2})
There is also one rare method, which adopts \textit{flashlight (B2.3)} patterns to indicate the performance of the model. For example, an app named SSR\cite{github_gallery} includes such patterns when the confidence of classification prediction reaches 90\%. 

\underline{Aural (B3)} patterns provide verbal confirmation or indication to user actions.
This pattern can occur in different stages of AI feature usage.
One scenario is that some apps trigger an aural feedback (e.g., ``ka-chick'' sound) when users input something to the AI feature to confirm that the user action has been performed, and the input has been taken.
Another scenario is that some apps send audio feedback when AI feature completes the task to re-attract users' attention.
This is because the users may be distracted while waiting for the AI feature to process the input.
In addition, a third scenario is when the apps need some user inputs.
Specifically, the sound may be used to ask users to re-take the input while it fails to perform as the input is invalid, or require users to give a further additional input.
As an example, \texttt{Freja} \cite{github_gallery} offers an ID card verification feature, and it requires the user to give both the front and the back view of the ID card.
The app will give a sound after users take the first photo to remind them to take a second one. 
All these sounds are short to draw users' attention, and the apps will always accompany with some text hints to reveal the app requirement. 


\underline{Haptic (B4)} patterns are similar to the \textit{aural patterns (B2)}, and they can also be used to give confirmation of user actions and task completion and to draw users' attention for further input, while in a vibration form.
It is normally considered as a secondary feedback and is used with other patterns, such as \textit{pop-up message (B1.1)} and \textit{instruction (B1.2 \& B2.2)}.
However, in our dataset, all these patterns (B4) are triggered when a failure occurs, either when users provide a wrong or invalid input the AI feature fails to process, or when the AI feature has other failures.


\textbf{Statistics:}
As seen, over half (58.50\% - 444/759) of the interaction patterns belong to general feedback, and are used by most apps (73.7\% - 188/255) in our dataset. 
Textual and graphical feedback (91.67\% - 257/444) are commonly used as they may be more intuitive to end-users.
In a comparison, auditory and haptic feedback only constitute 8.33\% (37/444), which may attribute to auditory and tactile sensory retardation.

\subsubsection{\textbf{Fine-grained feedback (C)}} \label{fg_feedback}
is defined as an advanced form of communication in which it returns detailed and customized feedback based on user input. 
It contains three different forms to present information to users, including textual, graphical and aural ways. 

\begin{figure*}
\centering
\subfloat[Text suggestion (C1.1)\protect\footnotemark]{\label{fig:C1.1} \includegraphics[width=0.20\textwidth]{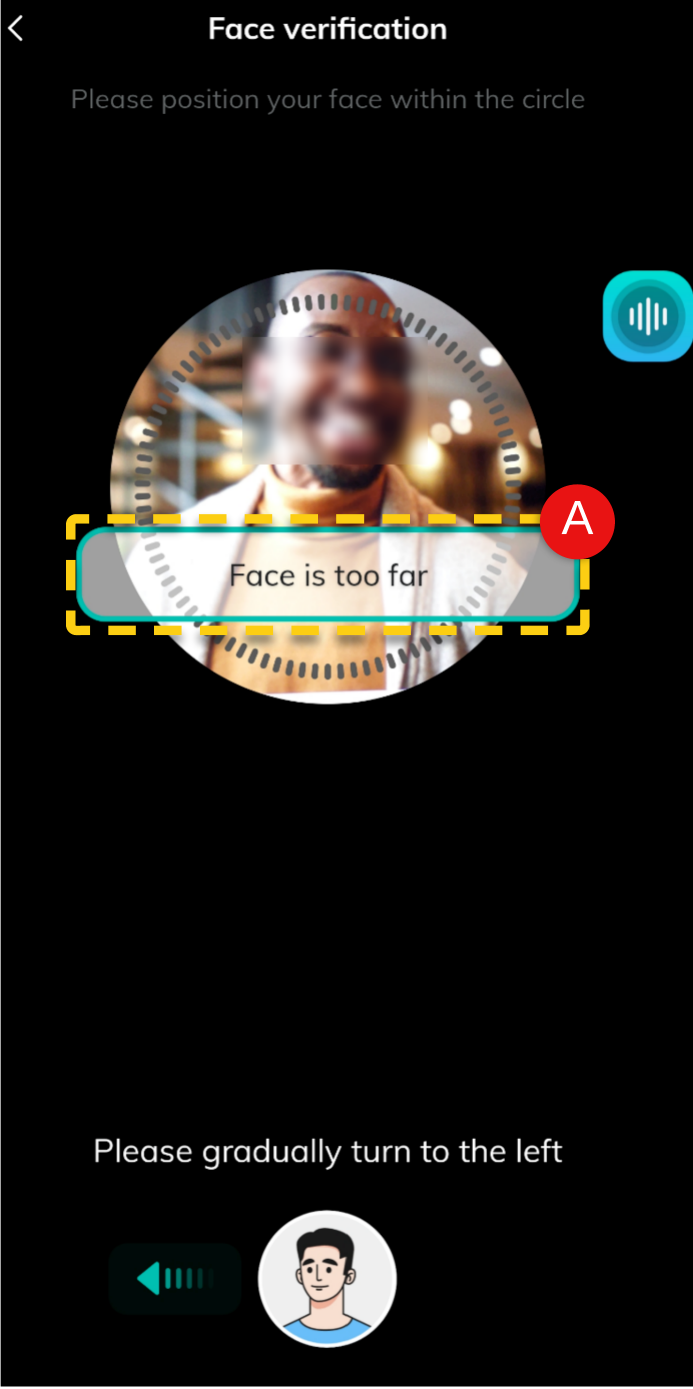}}%
\hfill
\subfloat[Object detector (C2.1).]{\label{fig:C2.1} \includegraphics[width=0.20 \textwidth]{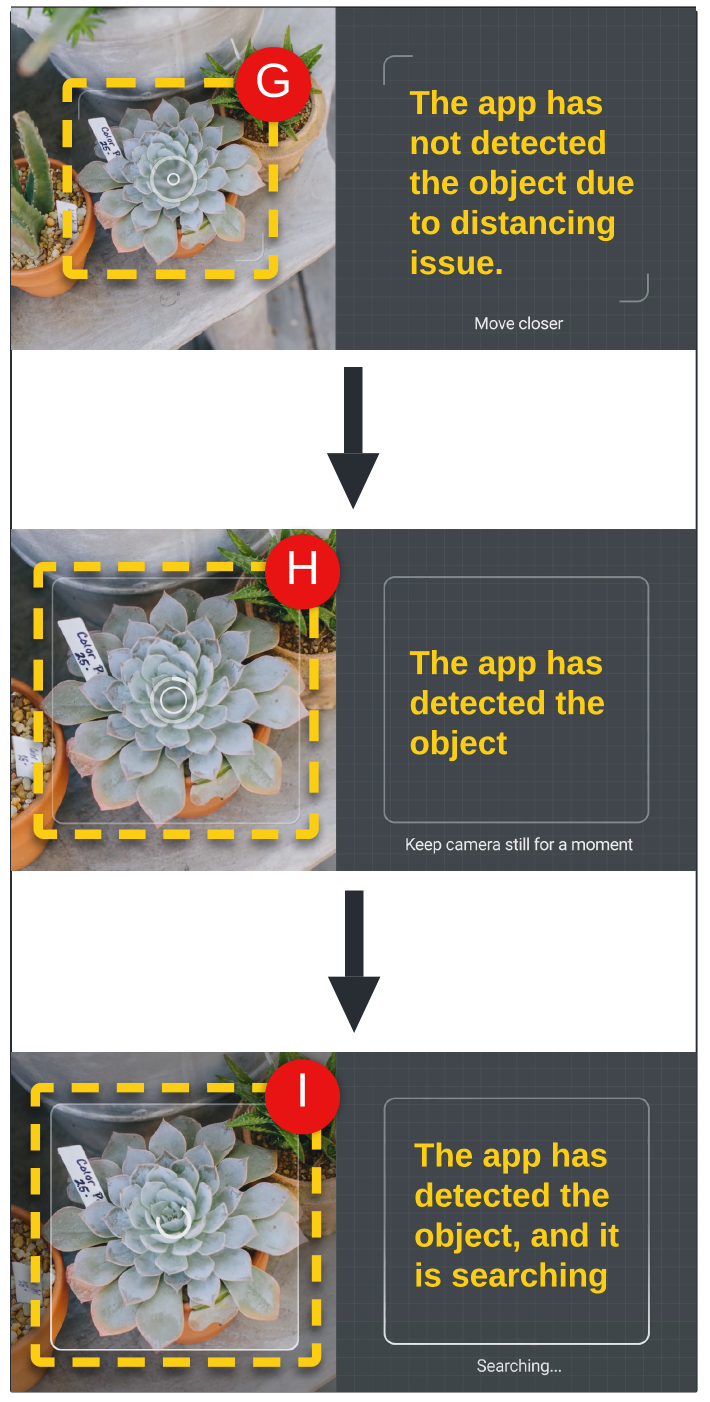}}%
\hfill
\subfloat[User personalisation by using the photo editing features (C2.2.1).]{\label{fig:C2.2.1} \includegraphics[width=0.20\textwidth]{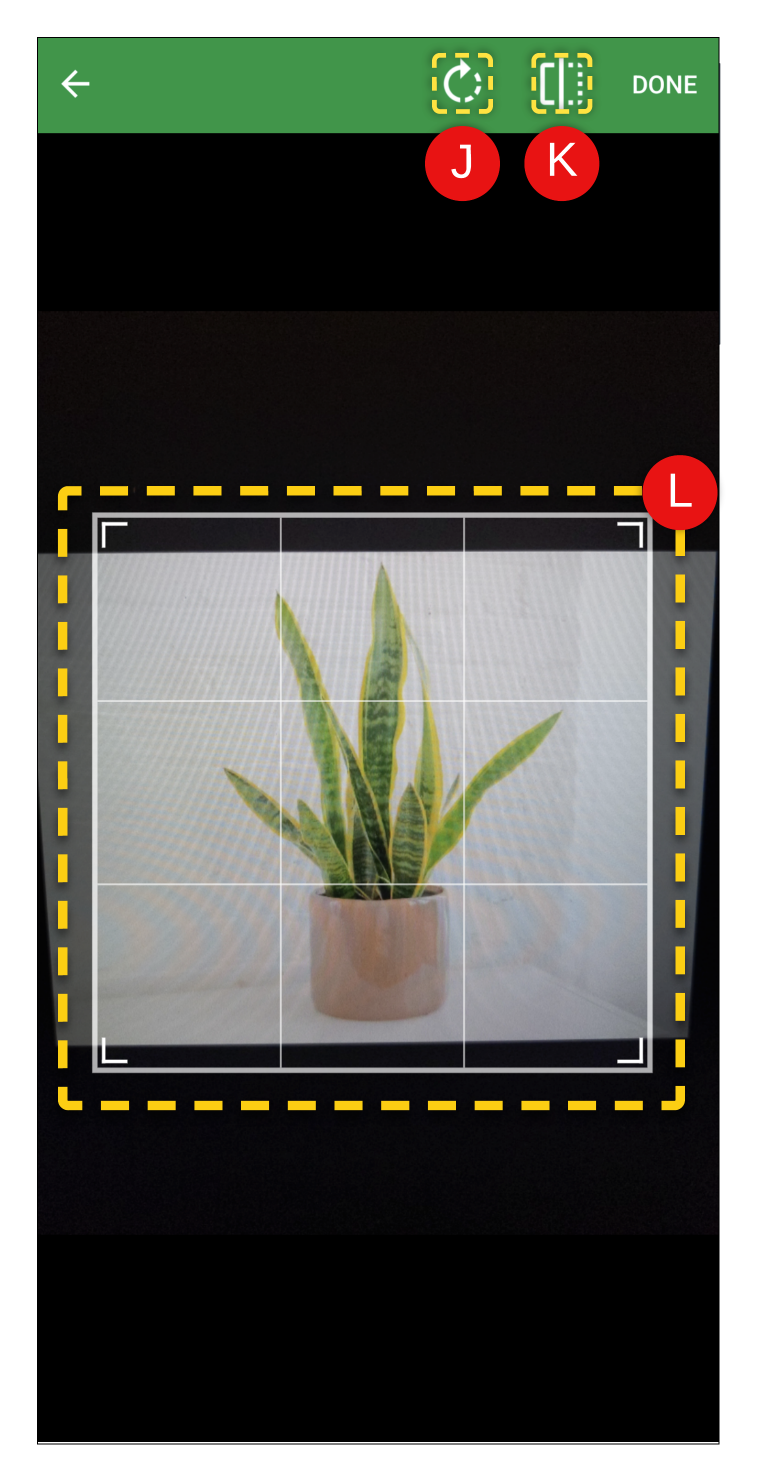}}%
\hfill
\subfloat[User personalisation by manually pointing the dots to the facial features (C2.2.2).]{\label{fig:C2.2.2} \includegraphics[width=0.20\textwidth]{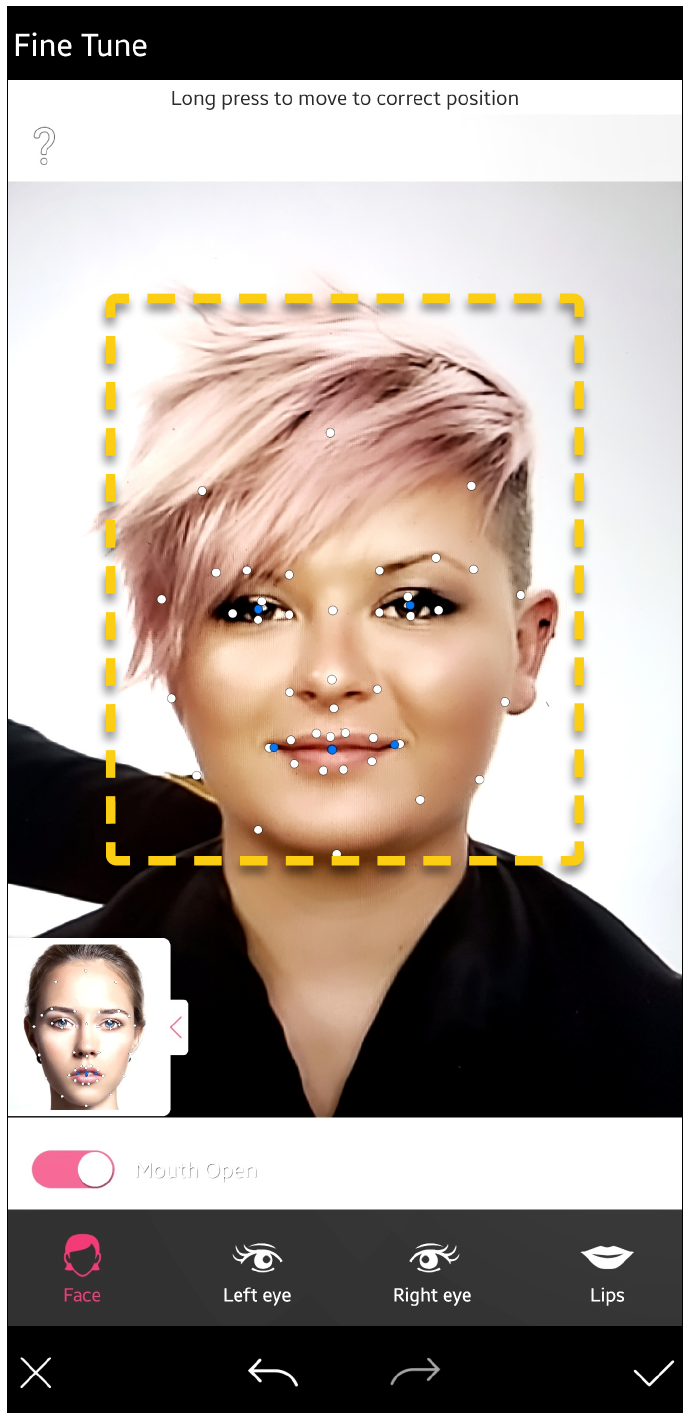}}%
\caption{Example of interaction pattern C.
A pattern shown in \raisebox{-0.5ex}{\includegraphics[width=3ex ,height=3 ex]{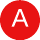}} is a text suggestion to remind user to put their face closer.
In pattern (C2.1), \raisebox{-0.5ex}{\includegraphics[width=3ex ,height=3 ex]{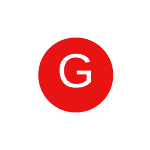}} shows that the shape of the frame is partial-rectangle, when the feature cannot detect the valid input. After finding the valid input, 
\raisebox{-0.5ex}{\includegraphics[width=3ex ,height=3 ex]{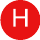}} illustrates that the shape is changed to be complete-rectangle; the frame in 
\raisebox{-0.5ex}{\includegraphics[width=3ex ,height=3 ex]{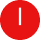}} is complete-rectangle with the background dimmed, as an indication that the feature is processing the input. In pattern (C.2.2.1), users could fine-tune the input before processing, with the feature of rotating shown in 
\raisebox{-0.5ex}{\includegraphics[width=3ex ,height=3 ex]{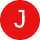}}, flipping shown in \raisebox{-0.5ex}{\includegraphics[width=3ex ,height=3 ex]{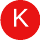}}, and cropping shown in \raisebox{-0.5ex}{\includegraphics[width=3ex ,height=3 ex]{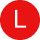}}
}
\Description{From left to right, Figure(a) shows an example of text suggestions (C1.1). The UI suggests "face is too far" to instruct users. The second example consists of three parts from top to bottom in the scenario where the user is trying to take a photo of a plant. When the app has not detected the object due to distancing issue, the UI shows an incomplete rectangle box to indicate the position of object. When the app detected the object, the dotted rectangle changed to a solid line. When the app starts searching, a loading icon will show in the middle of the screen. Figure (c) shows an example of user personalization patterns. The app allows users to use photo editing features, including rotating and flipping to change the input. Figure (d) shows another example of personalization feature. The UI shows a human face with several detection points around. Users can drag those points to adjust the beautification effect.}
\label{4figs2}
\end{figure*}
\footnotetext{We could not find the source of the face, so we blurred the face in (a) to protect privacy}

\begin{table*}
\centering
\setlength{\extrarowheight}{0pt}
\addtolength{\extrarowheight}{\aboverulesep}
\addtolength{\extrarowheight}{\belowrulesep}
\setlength{\aboverulesep}{0pt}
\setlength{\belowrulesep}{0pt}
\refstepcounter{table}
\label{diffOfTraditional_N_New_interaction}
\resizebox{1\linewidth}{!}{%
\begin{tabular}{clccccll} 
\toprule
\rowcolor[rgb]{0.902,0.902,0.902} \textbf{ID} & \textbf{Job title} & \textbf{Org.} & \textbf{Age} & \textbf{Gender} & \textbf{Experience} & \textbf{Educational background} & \textbf{\textbf{Experience}~in Al-related products} \\ 
\hline
P1 & \begin{tabular}[c]{@{}l@{}}Senior Experience\\ Designer\end{tabular} & IBM & 31-35 & M & Over 10 years & \begin{tabular}[c]{@{}l@{}}Business Information System; \\Computer Science\end{tabular} & NLP-related product for predictive forecasting\\ 
\hline
P2 & UX Lead & Google & 36-40 & M & Over 10 years & \begin{tabular}[c]{@{}l@{}}Human-Computer Interaction; Design, \\Computer Science;Business Information System;\\ Engineering\end{tabular} & ChromeOS; Google Assistant \\ 
\hline
P3 & Product Designer & Atlassian & 31-35 & F & 5-10 years & Business and Marketing & \begin{tabular}[c]{@{}l@{}}Media recommendation engine; \\Chatbot in a banking platform\end{tabular} \\ 
\hline
P4 & UX/UI designer & PwC & 31-35 & M & 3-5 years & Information Technology & No experience \\ 
\hline
P5 & UX Designer & Huawei & 26-30 & F & 5-10 years & Human Computer Interaction; Design & \begin{tabular}[c]{@{}l@{}}Huawei's personal voice assistant on mobile apps;\\Other products on SaaS platforms\end{tabular} \\ 
\hline
P6 & \begin{tabular}[c]{@{}l@{}}Senior Product \\Designer\end{tabular} & Microsoft & 26-30 & M & Over 10 years & Design & No experience \\
\bottomrule
\end{tabular}
}
\caption{Participant information}
\vspace{-7mm}
\label{tab:participants}
\end{table*}

\underline{Textual (C1)} are textual hints to help users adjust their input accordingly.
A banking app shown in Fig \ref{fig:C1.1} helps illustrate this pattern. It takes as input the user's face to verify the identity, so the user should put their face within the circular frame via a live camera. 
When the user does not take it properly, a text suggestion reminds the user \textit{``Face is too far."} for distancing issues.
Compared with the text form in general feedback (Type B), fine-grained feedback differs in the timing and the text content of the feedback.
In detail, the apps will decide which text hints to give users based on user input.
Providing specific instructions allows users to make an instant adjustment. Therefore, this feedback not only ensures the correct input to the AI features, but also helps engage with users.

\underline{Graphical (C2)} patterns are defined as using some graphical objects of different attributes to indicate the stages of AI features or to assist users in giving a valid input.
We identified three sub-types in our dataset.

\textit{Object detector (C2.1)} patterns are defined as using a dynamic camera frame to confirm that the app has detected the valid input, either by \textit{fixing the current frame (C2.1.1)} or \textit{changing the line style of the camera frame (C2.1.2)}.
These patterns are normally adopted when users are trying to take input via a camera.
For the first subtype (C2.1.1), once the app detects there is a target object in the current frame, it will fix it and avoid being further changed by user behaviours. 
Such fixing patterns can assist users better capture the object.
For the second type (C2.1.2), it is accompanied by the \textit{camera frame patterns (B2.1)} with a change in the line style of the camera frame once apps capture the target.
As seen in Fig~\ref{fig:C2.1}, when the app captures the plant, the line of the camera frame changes from a partial rectangular border to a continuous border.
\textit{Personalization (C2.2)} patterns allow users to choose the AI feature they would like to use and the level of modification they want.
Such patterns could give users a sense of control and let them feel more satisfied with the experience~\cite{nng_user_freedom}. 
One example is a photo editor (Fig~\ref{fig:C2.2.1} - C2.2.1). 
Users can crop and rotate their input picture before feeding to AI models to suit the input requirement.
Another example specifically caters to the facial-beatification apps where a user can control and instruct how AI should perform in terms of change area and the level of changing intensity by manually dragging the pointers to one's facial features such as eyes and lip (see Fig.~\ref{fig:C2.2.2} - C2.2.2). 



\textit{A audio spectrum visualiser (C2.3)} is a visual representation of a sound signal that indicates the real-time volume (intensity) and the frequency of user voice input. 
There are several forms of visualization, like soundwave and dribble forms.
A wavier and stronger animation means the sound is louder and quicker.

\underline{A fine-grained aural feedback (C3)} pattern synthesises the human voice to give different and richer feedback to users.
It is different from the \textit{general aural feedback (B4)} as it will react differently and semantically given different user inputs.
One interesting example is GlassesOn \cite{github_gallery} which is a medical app that measures a pupillary distance. 
This is achieved by asking a user to hold a magnetic card against one’s forehead and take a selfie. 
During this process, the app speaks to the user with some suggestions, like ``Your phone is held too low. Lift it to eye level'' and ``Your phone is tilted''. 
This kind of interaction is typically useful for apps deploying complex functionalities. 
This pattern can reduce the cognitive burden and simplify some complicated tasks.
App designers will often use equivalent text hints (B1 and C1) to assist the user as well.


\textbf{Statistics:}
With stronger capabilities to provide customized and instant feedback to users, 28.2\% (214/759) interaction patterns fall in the fine-grained feedback type, and 142/255 (55.7\%) AI features utilise these patterns.
Graphical feedback dominates the usage of this type, with 183/214 (85.5\%) instances.
Besides, textual feedback and auditory feedback have 17 and 14 implementations respectively.

\subsubsection{\textbf{Combination usage}}
\label{sec:RQ2_statistics}

During the process of forming our taxonomy, we recognized that many interaction patterns are used together to provide a better user experience. Among \nModelUsageTotal{} AI features from \nApp{} apps, 72/\nModelUsageTotal{} (\fpeval{round((72/\nModelUsageTotal{}),3) * 100}\%) of AI features include one interaction pattern, 166/\nModelUsageTotal{} (\fpeval{round((166/\nModelUsageTotal{}),3) * 100}\%) with 2 to 5, and 17/\nModelUsageTotal{} (\fpeval{round((17/\nModelUsageTotal{}),3) * 100}\%) include equal to or over 6 patterns. 
The most frequent combination is an instruction with graphical (B2.2) and textual (B1.2) elements (61/\nModelUsageTotal{} - \fpeval{round((61/\nModelUsageTotal{}),3) * 100}\%).
GlassOn \cite{github_gallery}, a medical app to measure users' pupillary distance, uses the most patterns (N=7), including \textit{pop-up messages (B1.1)},  \textit{textual and graphical instructions (B1.2 and B2.2)},  \textit{vibration (B4)}, \textit{text suggestion(C1.1)}, \textit{object detector (C2.2)} and \textit{fine-grained audio feedback (C3)}.
This is to ensure users understand the special requirements (e.g., holding a magnetic card against the forehead while taking selfie) and take correct inputs.

Another example is an app, SSR, which supports an audio classification feature. 
Apart from showing the types of input sounds, it implements a \textit{pop-up message (B1.1)}, \textit{flashlight (B2.3)}, \textit{auditory feedback (B3)} and \textit{vibration (B4)}  to inform the user that the prediction is above 90\% confidence. B1.1 is the primary interaction way while the rest are served as the secondary interactions. 
Combining different feedbacks helps conveying messages efficiently, and it also provides an engaging experience to end-users and enhances the transparency of AI features.

\section{Gallery Implementation}
To display our findings to practitioners, we organized our observations into an interaction gallery\cite{gallery} that introduces the usage and purpose of our research and showcases the findings.
It supports multi-faceted search by different attributes, including app category, model usage, and the types of patterns, for users to find concrete real-world examples of each interaction pattern. 
Such a gallery may be used to assist designers in mitigating the challenges identified in Section~\ref{sec:motivation}. We also leveraged this tool to conduct a user study to verify the usefulness of our work in Section~\ref{sec:usefulness}.
We hosted the website on GitHub, and all data and code are also stored in the same GitHub repository.
Note that it is just a research prototype to show our findings, more engineering efforts are needed to make it a practical tool for the community.

\section{Usefulness Evaluation}
\label{sec:usefulness}

To understand the usefulness and applicability of our findings and refine our taxonomy, we conducted a user study with industrial practitioners.


\subsection{Experimental Setup}
\label{sec:user_study_setup_rq3}



\subsubsection{Participants.}
By leveraging our network, we recruited six professional designers (2 female and 4 male) as our participants in this study.
Table~\ref{tab:participants} reports their background information and the types of AI-related products they had worked on before.
Participants spanned ages 26-40 with 2 in the age range of 26-30, 3 aged 31-35, and 1 aged 36-40. 
The participants' experience in working in UI/UX interaction design varies as follow: 3-4 years (1 participant), 5-9 years (2 participants) and 10+ years (3 participants).
Four of them have designed AI features before, and two did not have.
Of these participants, 3 have an educational background in Computer Science or Information Technology, 3 only have HCI, Design or Business educational background. 

\subsubsection{Study procedures.}
We arranged an interview with each participant.
Each interview lasted approximately 60 minutes. It was conducted remotely using a video conferencing tool, Zoom \footnote{https://zoom.us}, and we asked for the participants' consent to record the interview for accurate transcription and assisting further analysis. 
Each participant was compensated with a Gift Card worth \$100 for their participation.

The overall process in the interview is comprised of five steps. 
We started by briefing participants about the aim of our study and showing them some existing AI-powered apps (e.g., Snapchat - face detection feature) to get them involved (Step 1).
We also introduced the importance of feedback from apps to user input, 
and listed several examples to let them better understand our study.
Participants were then asked to talk about their previous interaction designs of AI-related apps, to see if they understand our goal. 
After that, we provided one task and asked them to think aloud their interaction design given the task (Step 2).
For each task, we elaborated on the functionalities of the AI feature, the input modality and its usage. 
We also provided some illustrative examples of app behaviours when the user inputs valid and invalid data to remind them to think about ensuring the quality of input. 

Later, we showcased our findings and the gallery we developed to the participants (Step 3), and asked them to refine the interaction design for the same task using the tool (Step 4).
We briefly gave a short demonstration of the gallery to avoid potential irrelevant issues about the web design of the gallery. 
In Step 5, we asked the participants about the reasons that they refined their first design to understand the usage of our taxonomy and gallery and the challenges they met.
After that, the participants were required to rate the usefulness of our taxonomy and collected examples shown in the gallery using a 5-point Likert scale in terms of difficulty in understanding the taxonomy, comprehensiveness and usefulness of the taxonomy and gallery. 
For each rating, the participants were asked to provide some reasons.
Finally, the participants were encouraged to give some general comments in terms of the taxonomy, the gallery or any other thoughts they have.

\subsubsection{Selected tasks.}
To ensure the diversity of the tasks, we selected and designed different tasks with different levels of complexity and popularity based on our findings in Section~\ref{sec:findings}. 
As seen in Fig~\ref{tab:task_user_study}, we chose a face detection task, a voice searching task and a pose detection task \footnotetext{The detailed description and results of each task can be seen in supplementary materials.}.
These tasks take different modalities, i.e., image or audio, as input.
In terms of complexity and popularity, the face detection features (41.6\%) are the most widely used and easy-to-understand while the pose detection task (1\%) is quite novel with a small number of examples in our gallery. The complexity and popularity of voice searching task falls in between. 
Each task will be assigned to two different participants, and each participant only needs to finish one task. 


\begin{table}
    \centering
    \setlength{\extrarowheight}{0pt}
    \addtolength{\extrarowheight}{\aboverulesep}
    \addtolength{\extrarowheight}{\belowrulesep}
    \setlength{\aboverulesep}{0pt}
    \setlength{\belowrulesep}{0pt}
    \resizebox{1\linewidth}{!}{%
    \begin{tabular}{c l l p{0.4\linewidth}} 
    \toprule
    \rowcolor[rgb]{0.902,0.902,0.902} \multicolumn{1}{l}{\textbf{Task}} & \textbf{Input modality} & \textbf{AI feature} & \textbf{Description}  \\ 
    \hline
    1 & Vision & Facial make-up & An AI feature auto-detects
users’ face and add cute motion stickers or refine their facial features.  \\ 
    \hline
    2 & Audio & Voice searching & An AI feature converts
the voice input into text for searching.  \\ 
    \hline
    3 & Vision & Body posture checking & An AI feature
detects the user's body to check if the sitting posture is correct. \\
    \bottomrule
    \end{tabular}
    }
    \caption[Caption for LOF]{Tasks in the user study}
   \label{tab:task_user_study}
   \vspace{-7mm}
\end{table}

\subsection{Results}
\label{sec:RQ3}

\subsubsection{Usefulness.}
Among all participants, five rated 4 or 5 and one gave a 3 score in terms of the usefulness of our findings.
They believe these terminology and examples cover apps with different AI features and different interaction patterns, from which they can draw some inspirations and refine their design.
P4 commented \textit{``The gallery comes from the actual designs from actual apps. It may not look that fancy but it is real so I think the inspiration from what's real is more valuable.''}


In detail, they believe our terminology and gallery \cite{gallery} is a new and novel tool that can support interaction design for AI features, as the existing design sharing platforms (e.g., Dribbble, Muzli, Mobbin)
or guidelines~\cite{material_io, ios_design} mainly focus on general apps or give high-level guidance. Their thoughts also aligns well with our findings in Section~\ref{sec:motivation}.
While some existing design guidelines~\cite{microsoft, apple_coreML}  provide conceptual guidance in designing AI features, the participants think that our tool gives more detailed and concrete guidance of what and how to design for AI features so that they can easily balance the trade-off between the user-friendliness and simplicity of the design by the provided examples.

In addition, existing platforms like Product Hunt
provide some image-based examples from AI-powered apps, but fail to give a systematic evaluation of the underlying patterns.
Our taxonomy and gallery well-summarise the existing interaction patterns in AI features, and the participants find it easier to understand, apply and generalize to their own tasks.
P2 said: \textit{``As a designer, we're really open to a lot of different interpretations of our designs... Different implementations of the same pattern from different AI features and app categories could help me understand the way that how other products are design to capture user input... I would design a lot of different interactions inspired from(apps) under a different category rather than its own types of product''}

As the gallery allows users to search in terms of the patterns of AI features, interactions patterns and app categories, the participants found this multi-faceted search function aligned well with their working process when designing the interaction design.
The normal procedure is that they would first search for the designs of other apps with similar features (AI features), and then found out the specific components they want to implement and get inspired from others (interaction patterns). 
Both are supported by our findings and the gallery.
The video-based examples provided by our gallery are also intuitive and more natural than some image examples due to the nature of interaction.
They really appreciate our work and one of the participants believe our findings in mobile apps can be also generalized to other devices, like websites and PC apps.
They also ask if we will release the gallery in the future as they are willing to use our tool in their work.

\subsubsection{Comprehensiveness.}
In terms of comprehensiveness, most participants except P3 thought that the gallery covered enough examples for each pattern and scored 4.43/5 on average, albeit with the encouragement for adding more examples. 
In terms of whether the terms are easy-to-understand, three participants rated 3 and two rated 4. 
The participants said that while at the first glance, the terms are not easy to understand, they could 
They think the terms may not be easily understood at the first glance, but after seeing the illustrative examples, everything becomes clear and make sense.
P2 gave some suggestions to refine some terms to better fits with terms used by the designer community.
For example, we initially named Type C as ``dynamic feedback'', but the participant think Type B is also dynamic.
Therefore, we renamed these two types as general feedback and fine-grained feedback to better differentiate them.

\section{IMPLICATION}
\label{Sec:discussion}



\textbf{On the effectiveness of interaction design.} 
With the contributions of this paper, designers can take advantage of the concrete set of interaction patterns and their prevalence as an inspiration to improve their interaction design. 
However, since our study focuses on summarising the patterns, we do not evaluate the usefulness and effectiveness of different patterns.
Note that fine-grained feedback is not always better than minimum feedback, but depends on the detailed usage scenario.
A good example mentioned in delayed feedback (A3) is \texttt{Smart alarm} \cite{github_gallery}; the app intends not to interact with users instantly while sleeping. If a fine-grained feedback is used in this context, it not only adds complexity to the feature, more importantly, but also brings a bad user experience to users.  
In addition, the importance of AI features within the mobile app may also influence the selection of feedback patterns considering the balance of cost and benefit, especially for start-up teams.
More research is needed to understand the effectiveness of different feedback patterns in term of app category, feature type, importance, cost, etc. 


\textbf{On mitigating human error.} We found that the summarized patterns can be applied in mitigating human error. 
According to Norman \cite{nng_slips}, there are two kinds of user errors: mistakes and slips. 
Mistakes arise when users have inappropriate goals for the current task. 
For example, a mistake happens when a user misunderstands the face detection feature, and takes a picture without a face inside, leading to an unexpected outcome.
Whereas a slip occurs when the user properly understands the AI feature, but accidentally provides a flawed input because of incorrect execution (e.g., blurred photo, lighting issues).
An instructions with textual (B1.2) and graphical (B2.2) instruction can help avoid potential mistakes by ensuring users understand what the AI features can do and can not do.
While these two patterns could also help mitigate slip errors, patterns that correct users when they are using the AI features would be more effective and useful.
Such patterns include \textit{camera frame (B2.1)}, \textit{text suggestions (C1.1)}, \textit{graphical fine-grained feedback (C2.1, C2.2, C2.3)}, and \textit{auditory feedback (C3)}.
As for the content of the hints, Google \cite{g_PAIR} provides some guidance on this, such as a message on miscalibrated input needs to include an action and an explanation of that action. 
For instance, an app uses a pop-up message hinting \textit{``Move a camera closer to the object.''}.
In addition, the interaction design should also take underperformed users (e.g., the aged, and disabled) into consideration, to engage them into society.

\textbf{Potential Applications.}
We also asked the participants in our user study to discuss potential applications based on our taxonomy and gallery. 
They identified three applications that could be enabled by our work in different designing stages, from education and knowledge discovery to internal communication.

For the education application, they believe this tool can help students or early career designers to learn some knowledge before implementing interaction design to a real AI product.
For example, the universities or other organisations can offer some courses, leveraging the well-summarised terminology and corresponding examples, to teach students or train early career designers to identify the trade-off between each design pattern, and understand the usage scenarios, i.e., when (not) to adopt one pattern. Note that while our findings do not evaluate the quality or usefulness of each pattern and design, they reflect the popularity of different patterns.

Additionally, all participants agree that our work could also assist their work when they start prototyping an AI features in the knowledge discovery phase.
Not only could our work give them ideas and inspiration for how they can develop their own prototypes, but our findings also enable competitive analysis.
It can give them an overview of how existing apps (i.e., competitors) implemented the interaction design so that they can reference the effective ones or improve/avoid the ineffective ones.
Moreover, our findings save their resources in the part of product research and development (R\&D) as they do not need to manually find similar products, download and try to explore the interactions in each app. 
In particular, P5 said \textit{``In my organisation, we also do the recording or take screenshots of each source, all of which will be used for a discussion on the design and another idea. Your findings save our time and ease the process of design discovery.''}. P2 further elaborated on this point by saying that \textit{``Using your findings save time and money for research. We could allocate more resources on user testing, and potentially fast-track the launch of the minimum viable product (MVP).''}


The third case is about boosting internal communication. 
Previous works mentioned there is a knowledge gap between UX designers and AI engineers \cite{ux_design_as_ML_mat, investigating_yang, yang2018mapping}. This is also confirmed by our participants and they think that our findings can be a bridge to connect them with the engineers. Specifically, they can use our examples to communicate with AI engineers in a more intuitive and efficient way. By doing so, the engineer could also evaluate the hardness and possibility of implementing such designs.
Apart from this, participants also believe our work can assist the decision-making process of other stakeholders like the product manager. 
For example, after seeing other relevant products design, some questions like \textit{``are there any gaps to launch the new product with better UX? Does our team have enough capacity and resources to take on a project or a feature?''}, could be raised and answered.
Our findings can be used to evaluate the complexity of the project from the UX perspective and hence aid the process of stakeholders' decision-making.

\section{CONCLUSION and FUTURE WORK}

In this work, we identified three challenges in designing interaction for AI features, namely unfamiliarity to AI features for users, insufficient knowledge to have a proper expectation of AI features and insufficient tool or gallery for designing interaction of AI features. 
To address these issues, we conducted a systematic large-scale analysis on mobile apps to investigate the usage of AI features and the implemented interaction patterns. We also implemented our findings into a gallery to assist the following usage. The participants in our user study give positive feedback, which confirms the usefulness, generalizability and future applications of our work. 
The outcomes highlight that, when designing AI-related apps, the taxonomy is generalisable across different AI features, and our gallery is different from existing design sharing platform, in which the real-world examples inspire designers' research and save their resources on the stage of Resource and Development (R\&D). 

We hope that our work could be beneficial to design better and more human-centric AI apps, and facilitate further research. 
With the advent of AI models, we see significant value in further enriching and refining taxonomy for mobile human-AI interaction.
Moreover, we believe that our findings could be easily extended to iOS platform, and even to other devices.
In the future, as suggested by P2 in our user study, some potential future work could be on exploring interactions in AR/VR/MR (Augmented/Virtual/Mixed Reality) devices. 


\bibliographystyle{ACM-Reference-Format}
\bibliography{Bibliography.bib}


\begin{thebibliography}{49}


\ifx \showCODEN    \undefined \def \showCODEN     #1{\unskip}     \fi
\ifx \showDOI      \undefined \def \showDOI       #1{#1}\fi
\ifx \showISBNx    \undefined \def \showISBNx     #1{\unskip}     \fi
\ifx \showISBNxiii \undefined \def \showISBNxiii  #1{\unskip}     \fi
\ifx \showISSN     \undefined \def \showISSN      #1{\unskip}     \fi
\ifx \showLCCN     \undefined \def \showLCCN      #1{\unskip}     \fi
\ifx \shownote     \undefined \def \shownote      #1{#1}          \fi
\ifx \showarticletitle \undefined \def \showarticletitle #1{#1}   \fi
\ifx \showURL      \undefined \def \showURL       {\relax}        \fi
\providecommand\bibfield[2]{#2}
\providecommand\bibinfo[2]{#2}
\providecommand\natexlab[1]{#1}
\providecommand\showeprint[2][]{arXiv:#2}

\bibitem[Alharbi and Yeh(2015)]%
        {alharbi2015collect}
\bibfield{author}{\bibinfo{person}{Khalid Alharbi} {and} \bibinfo{person}{Tom
  Yeh}.} \bibinfo{year}{2015}\natexlab{}.
\newblock \showarticletitle{Collect, decompile, extract, stats, and diff:
  Mining design pattern changes in Android apps}. In
  \bibinfo{booktitle}{\emph{Proceedings of the 17th International Conference on
  Human-Computer Interaction with Mobile Devices and Services}}. ACM,
  \bibinfo{pages}{515--524}.
\newblock


\bibitem[Amershi et~al\mbox{.}(2019)]%
        {microsoft}
\bibfield{author}{\bibinfo{person}{Saleema Amershi}, \bibinfo{person}{Dan
  Weld}, \bibinfo{person}{Mihaela Vorvoreanu}, \bibinfo{person}{Adam Fourney},
  \bibinfo{person}{Besmira Nushi}, \bibinfo{person}{Penny Collisson},
  \bibinfo{person}{Jina Suh}, \bibinfo{person}{Shamsi Iqbal},
  \bibinfo{person}{Paul~N Bennett}, \bibinfo{person}{Kori Inkpen},
  {et~al\mbox{.}}} \bibinfo{year}{2019}\natexlab{}.
\newblock \showarticletitle{Guidelines for human-AI interaction}. In
  \bibinfo{booktitle}{\emph{Proceedings of the 2019 chi conference on human
  factors in computing systems}}. \bibinfo{pages}{1--13}.
\newblock


\bibitem[Author(s)(2022a)]%
        {github_gallery}
\bibfield{author}{\bibinfo{person}{Anonymous Author(s)}.}
  \bibinfo{year}{2022}\natexlab{a}.
\newblock \bibinfo{title}{Github. Interaction design gallery.}
\newblock
\newblock
\urldef\tempurl%
\url{https://github.com/algaeSpace/interaction_gallery}
\showURL{%
\tempurl}


\bibitem[Author(s)(2022b)]%
        {gallery}
\bibfield{author}{\bibinfo{person}{Anonymous Author(s)}.}
  \bibinfo{year}{2022}\natexlab{b}.
\newblock \bibinfo{title}{Interaction design gallery.}
\newblock
\newblock
\urldef\tempurl%
\url{https://algaespace.github.io/interaction_gallery/Gallery_static/index_page/search.html}
\showURL{%
\tempurl}


\bibitem[Coursaris and Kim(2011)]%
        {coursaris2011meta}
\bibfield{author}{\bibinfo{person}{Constantinos~K Coursaris} {and}
  \bibinfo{person}{Dan~J Kim}.} \bibinfo{year}{2011}\natexlab{}.
\newblock \showarticletitle{A meta-analytical review of empirical mobile
  usability studies}.
\newblock \bibinfo{journal}{\emph{Journal of usability studies}}
  \bibinfo{volume}{6}, \bibinfo{number}{3} (\bibinfo{year}{2011}),
  \bibinfo{pages}{117--171}.
\newblock


\bibitem[Deng et~al\mbox{.}(2022)]%
        {deng2022understanding}
\bibfield{author}{\bibinfo{person}{Zizhuang Deng}, \bibinfo{person}{Kai Chen},
  \bibinfo{person}{Guozhu Meng}, \bibinfo{person}{Xiaodong Zhang},
  \bibinfo{person}{Ke Xu}, {and} \bibinfo{person}{Yao Cheng}.}
  \bibinfo{year}{2022}\natexlab{}.
\newblock \showarticletitle{Understanding Real-world Threats to Deep Learning
  Models in Android Apps}. In \bibinfo{booktitle}{\emph{Proceedings of the 2022
  ACM SIGSAC Conference on Computer and Communications Security}}.
  \bibinfo{pages}{785--799}.
\newblock


\bibitem[Department(2022)]%
        {androidMarketShare}
\bibfield{author}{\bibinfo{person}{Statista~Research Department}.}
  \bibinfo{year}{2022}\natexlab{}.
\newblock \bibinfo{title}{Share of global smartphone shipments by operating
  system from 2014 to 2023}.
\newblock
\newblock
\urldef\tempurl%
\url{https://www.statista.com/statistics/272307/market-share-forecast-for-smartphone-operating-systems/#:~:text=Smartphones\%20running\%20the\%20Android\%20operating,percent\%20share\%20of\%20the\%20market.}
\showURL{%
\tempurl}


\bibitem[Devlin et~al\mbox{.}(2018)]%
        {devlin2018bert}
\bibfield{author}{\bibinfo{person}{Jacob Devlin}, \bibinfo{person}{Ming-Wei
  Chang}, \bibinfo{person}{Kenton Lee}, {and} \bibinfo{person}{Kristina
  Toutanova}.} \bibinfo{year}{2018}\natexlab{}.
\newblock \showarticletitle{Bert: Pre-training of deep bidirectional
  transformers for language understanding}.
\newblock \bibinfo{journal}{\emph{arXiv preprint arXiv:1810.04805}}
  (\bibinfo{year}{2018}).
\newblock


\bibitem[Dhar et~al\mbox{.}(2021)]%
        {dhar2021survey}
\bibfield{author}{\bibinfo{person}{Sauptik Dhar}, \bibinfo{person}{Junyao Guo},
  \bibinfo{person}{Jiayi Liu}, \bibinfo{person}{Samarth Tripathi},
  \bibinfo{person}{Unmesh Kurup}, {and} \bibinfo{person}{Mohak Shah}.}
  \bibinfo{year}{2021}\natexlab{}.
\newblock \showarticletitle{A survey of on-device machine learning: An
  algorithms and learning theory perspective}.
\newblock \bibinfo{journal}{\emph{ACM Transactions on Internet of Things}}
  \bibinfo{volume}{2}, \bibinfo{number}{3} (\bibinfo{year}{2021}),
  \bibinfo{pages}{1--49}.
\newblock


\bibitem[Dove et~al\mbox{.}(2017)]%
        {ux_design_as_ML_mat}
\bibfield{author}{\bibinfo{person}{Graham Dove}, \bibinfo{person}{Kim Halskov},
  \bibinfo{person}{Jodi Forlizzi}, {and} \bibinfo{person}{John Zimmerman}.}
  \bibinfo{year}{2017}\natexlab{}.
\newblock \showarticletitle{UX design innovation: Challenges for working with
  machine learning as a design material}. In
  \bibinfo{booktitle}{\emph{Proceedings of the 2017 chi conference on human
  factors in computing systems}}. \bibinfo{pages}{278--288}.
\newblock


\bibitem[Fu et~al\mbox{.}(2013)]%
        {fu2013people}
\bibfield{author}{\bibinfo{person}{Bin Fu}, \bibinfo{person}{Jialiu Lin},
  \bibinfo{person}{Lei Li}, \bibinfo{person}{Christos Faloutsos},
  \bibinfo{person}{Jason Hong}, {and} \bibinfo{person}{Norman Sadeh}.}
  \bibinfo{year}{2013}\natexlab{}.
\newblock \showarticletitle{Why people hate your app: Making sense of user
  feedback in a mobile app store}. In \bibinfo{booktitle}{\emph{Proceedings of
  the 19th ACM SIGKDD international conference on Knowledge discovery and data
  mining}}. ACM, \bibinfo{pages}{1276--1284}.
\newblock


\bibitem[Gong et~al\mbox{.}(2004)]%
        {handheld_mobile_device_guidelines}
\bibfield{author}{\bibinfo{person}{Jun Gong}, \bibinfo{person}{Peter
  Tarasewich}, {et~al\mbox{.}}} \bibinfo{year}{2004}\natexlab{}.
\newblock \showarticletitle{Guidelines for handheld mobile device interface
  design}. In \bibinfo{booktitle}{\emph{Proceedings of DSI 2004 Annual
  Meeting}}. Citeseer, \bibinfo{pages}{3751--3756}.
\newblock


\bibitem[Google(2020)]%
        {M_IO_ML}
\bibfield{author}{\bibinfo{person}{Google}.} \bibinfo{year}{2020}\natexlab{}.
\newblock \bibinfo{title}{Machine learning patterns}.
\newblock
\newblock
\urldef\tempurl%
\url{https://m2.material.io/design/machine-learning/understanding-ml-patterns.html#}
\showURL{%
\tempurl}


\bibitem[Google(2022a)]%
        {g_PAIR}
\bibfield{author}{\bibinfo{person}{Google}.} \bibinfo{year}{2022}\natexlab{a}.
\newblock \bibinfo{title}{Google - People + AI Research (PAIR)}.
\newblock
\newblock
\urldef\tempurl%
\url{https://pair.withgoogle.com/}
\showURL{%
\tempurl}


\bibitem[Google(2022b)]%
        {ml-kit}
\bibfield{author}{\bibinfo{person}{Google}.} \bibinfo{year}{2022}\natexlab{b}.
\newblock \bibinfo{title}{Machine learning for mobile developers}.
\newblock
\newblock
\urldef\tempurl%
\url{https://developers.google.com/ml-kit}
\showURL{%
\tempurl}


\bibitem[Google(2022c)]%
        {material_io}
\bibfield{author}{\bibinfo{person}{Google}.} \bibinfo{year}{2022}\natexlab{c}.
\newblock \bibinfo{title}{Material Design}.
\newblock
\newblock
\urldef\tempurl%
\url{https://material.io}
\showURL{%
\tempurl}


\bibitem[Google(2022d)]%
        {tfHub_domain}
\bibfield{author}{\bibinfo{person}{Google}.} \bibinfo{year}{2022}\natexlab{d}.
\newblock \bibinfo{title}{TensorFlow Hub}.
\newblock
\newblock
\urldef\tempurl%
\url{https://tfhub.dev/}
\showURL{%
\tempurl}


\bibitem[Gulati et~al\mbox{.}(2020)]%
        {gulati2020conformer}
\bibfield{author}{\bibinfo{person}{Anmol Gulati}, \bibinfo{person}{James Qin},
  \bibinfo{person}{Chung-Cheng Chiu}, \bibinfo{person}{Niki Parmar},
  \bibinfo{person}{Yu Zhang}, \bibinfo{person}{Jiahui Yu}, \bibinfo{person}{Wei
  Han}, \bibinfo{person}{Shibo Wang}, \bibinfo{person}{Zhengdong Zhang},
  \bibinfo{person}{Yonghui Wu}, {et~al\mbox{.}}}
  \bibinfo{year}{2020}\natexlab{}.
\newblock \showarticletitle{Conformer: Convolution-augmented transformer for
  speech recognition}.
\newblock \bibinfo{journal}{\emph{arXiv preprint arXiv:2005.08100}}
  (\bibinfo{year}{2020}).
\newblock


\bibitem[Holmquist(2017)]%
        {holmquist2017intelligence}
\bibfield{author}{\bibinfo{person}{Lars~Erik Holmquist}.}
  \bibinfo{year}{2017}\natexlab{}.
\newblock \showarticletitle{Intelligence on tap: artificial intelligence as a
  new design material}.
\newblock \bibinfo{journal}{\emph{interactions}} \bibinfo{volume}{24},
  \bibinfo{number}{4} (\bibinfo{year}{2017}), \bibinfo{pages}{28--33}.
\newblock


\bibitem[Howard et~al\mbox{.}(2017)]%
        {howard2017mobilenets}
\bibfield{author}{\bibinfo{person}{Andrew~G Howard}, \bibinfo{person}{Menglong
  Zhu}, \bibinfo{person}{Bo Chen}, \bibinfo{person}{Dmitry Kalenichenko},
  \bibinfo{person}{Weijun Wang}, \bibinfo{person}{Tobias Weyand},
  \bibinfo{person}{Marco Andreetto}, {and} \bibinfo{person}{Hartwig Adam}.}
  \bibinfo{year}{2017}\natexlab{}.
\newblock \showarticletitle{Mobilenets: Efficient convolutional neural networks
  for mobile vision applications}.
\newblock \bibinfo{journal}{\emph{arXiv preprint arXiv:1704.04861}}
  (\bibinfo{year}{2017}).
\newblock


\bibitem[Huang and Chen(2022)]%
        {huang2022smart}
\bibfield{author}{\bibinfo{person}{Yujin Huang} {and} \bibinfo{person}{Chunyang
  Chen}.} \bibinfo{year}{2022}\natexlab{}.
\newblock \showarticletitle{Smart app attack: hacking deep learning models in
  android apps}.
\newblock \bibinfo{journal}{\emph{IEEE Transactions on Information Forensics
  and Security}}  \bibinfo{volume}{17} (\bibinfo{year}{2022}),
  \bibinfo{pages}{1827--1840}.
\newblock


\bibitem[Huang et~al\mbox{.}(2021)]%
        {huang2021robustness}
\bibfield{author}{\bibinfo{person}{Yujin Huang}, \bibinfo{person}{Han Hu},
  {and} \bibinfo{person}{Chunyang Chen}.} \bibinfo{year}{2021}\natexlab{}.
\newblock \showarticletitle{Robustness of on-device models: Adversarial attack
  to deep learning models on android apps}. In \bibinfo{booktitle}{\emph{2021
  IEEE/ACM 43rd International Conference on Software Engineering: Software
  Engineering in Practice (ICSE-SEIP)}}. IEEE, \bibinfo{pages}{101--110}.
\newblock


\bibitem[Iandola et~al\mbox{.}(2016)]%
        {iandola2016squeezenet}
\bibfield{author}{\bibinfo{person}{Forrest~N Iandola}, \bibinfo{person}{Song
  Han}, \bibinfo{person}{Matthew~W Moskewicz}, \bibinfo{person}{Khalid Ashraf},
  \bibinfo{person}{William~J Dally}, {and} \bibinfo{person}{Kurt Keutzer}.}
  \bibinfo{year}{2016}\natexlab{}.
\newblock \showarticletitle{SqueezeNet: AlexNet-level accuracy with 50x fewer
  parameters and< 0.5 MB model size}.
\newblock \bibinfo{journal}{\emph{arXiv preprint arXiv:1602.07360}}
  (\bibinfo{year}{2016}).
\newblock


\bibitem[Inc.(2022a)]%
        {apple_coreML}
\bibfield{author}{\bibinfo{person}{Apple Inc.}}
  \bibinfo{year}{2022}\natexlab{a}.
\newblock \bibinfo{title}{Core ML - Machine Learning}.
\newblock
\newblock
\urldef\tempurl%
\url{https://developer.apple.com/machine-learning/core-ml}
\showURL{%
\tempurl}


\bibitem[Inc.(2022b)]%
        {ios_design}
\bibfield{author}{\bibinfo{person}{Apple Inc.}}
  \bibinfo{year}{2022}\natexlab{b}.
\newblock \bibinfo{title}{Designing for iOS}.
\newblock
\newblock
\urldef\tempurl%
\url{https://developer.apple.com/design/human-interface-guidelines/platforms/designing-for-ios/}
\showURL{%
\tempurl}


\bibitem[Inc.(2022c)]%
        {apple_ML_gl}
\bibfield{author}{\bibinfo{person}{Apple Inc.}}
  \bibinfo{year}{2022}\natexlab{c}.
\newblock \bibinfo{title}{Human Interface Guidelines - Machine Learning}.
\newblock
\newblock
\urldef\tempurl%
\url{https://developer.apple.com/design/human-interface-guidelines/technologies/machine-learning/introduction/}
\showURL{%
\tempurl}


\bibitem[Jahanian et~al\mbox{.}(2017a)]%
        {jahanian2017mining}
\bibfield{author}{\bibinfo{person}{Ali Jahanian}, \bibinfo{person}{Phillip
  Isola}, {and} \bibinfo{person}{Donglai Wei}.}
  \bibinfo{year}{2017}\natexlab{a}.
\newblock \showarticletitle{Mining Visual Evolution in 21 Years of Web Design}.
  In \bibinfo{booktitle}{\emph{Proceedings of the 2017 CHI Conference Extended
  Abstracts on Human Factors in Computing Systems}}. ACM,
  \bibinfo{pages}{2676--2682}.
\newblock


\bibitem[Jahanian et~al\mbox{.}(2017b)]%
        {jahanian2017colors}
\bibfield{author}{\bibinfo{person}{Ali Jahanian}, \bibinfo{person}{Shaiyan
  Keshvari}, \bibinfo{person}{SVN Vishwanathan}, {and} \bibinfo{person}{Jan~P
  Allebach}.} \bibinfo{year}{2017}\natexlab{b}.
\newblock \showarticletitle{Colors--Messengers of Concepts: Visual Design
  Mining for Learning Color Semantics}.
\newblock \bibinfo{journal}{\emph{ACM Transactions on Computer-Human
  Interaction(TOCHI)}} \bibinfo{volume}{24}, \bibinfo{number}{1}
  (\bibinfo{year}{2017}).
\newblock


\bibitem[Laubheimer(2015)]%
        {nng_slips}
\bibfield{author}{\bibinfo{person}{Page Laubheimer}.}
  \bibinfo{year}{2015}\natexlab{}.
\newblock \bibinfo{title}{Preventing User Errors: Avoiding Unconscious Slips}.
\newblock
\newblock
\urldef\tempurl%
\url{https://www.nngroup.com/articles/slips/}
\showURL{%
\tempurl}


\bibitem[LeCun et~al\mbox{.}(2015)]%
        {lecun2015deep}
\bibfield{author}{\bibinfo{person}{Yann LeCun}, \bibinfo{person}{Yoshua
  Bengio}, {and} \bibinfo{person}{Geoffrey Hinton}.}
  \bibinfo{year}{2015}\natexlab{}.
\newblock \showarticletitle{Deep learning}.
\newblock \bibinfo{journal}{\emph{nature}} \bibinfo{volume}{521},
  \bibinfo{number}{7553} (\bibinfo{year}{2015}), \bibinfo{pages}{436--444}.
\newblock


\bibitem[Liao et~al\mbox{.}(2020)]%
        {liao2020questioning}
\bibfield{author}{\bibinfo{person}{Q~Vera Liao}, \bibinfo{person}{Daniel
  Gruen}, {and} \bibinfo{person}{Sarah Miller}.}
  \bibinfo{year}{2020}\natexlab{}.
\newblock \showarticletitle{Questioning the AI: informing design practices for
  explainable AI user experiences}. In \bibinfo{booktitle}{\emph{Proceedings of
  the 2020 CHI Conference on Human Factors in Computing Systems}}.
  \bibinfo{pages}{1--15}.
\newblock


\bibitem[Liao et~al\mbox{.}(2021)]%
        {IBM_question_driven}
\bibfield{author}{\bibinfo{person}{Q~Vera Liao}, \bibinfo{person}{Milena
  Pribi{\'c}}, \bibinfo{person}{Jaesik Han}, \bibinfo{person}{Sarah Miller},
  {and} \bibinfo{person}{Daby Sow}.} \bibinfo{year}{2021}\natexlab{}.
\newblock \showarticletitle{Question-Driven Design Process for Explainable AI
  User Experiences}.
\newblock \bibinfo{journal}{\emph{arXiv preprint arXiv:2104.03483}}
  (\bibinfo{year}{2021}).
\newblock


\bibitem[Lowe(1999)]%
        {lowe1999object}
\bibfield{author}{\bibinfo{person}{David~G Lowe}.}
  \bibinfo{year}{1999}\natexlab{}.
\newblock \showarticletitle{Object recognition from local scale-invariant
  features}. In \bibinfo{booktitle}{\emph{Proceedings of the seventh IEEE
  international conference on computer vision}}, Vol.~\bibinfo{volume}{2}.
  Ieee, \bibinfo{pages}{1150--1157}.
\newblock


\bibitem[Martin et~al\mbox{.}(2017)]%
        {martin2017survey}
\bibfield{author}{\bibinfo{person}{William Martin}, \bibinfo{person}{Federica
  Sarro}, \bibinfo{person}{Yue Jia}, \bibinfo{person}{Yuanyuan Zhang}, {and}
  \bibinfo{person}{Mark Harman}.} \bibinfo{year}{2017}\natexlab{}.
\newblock \showarticletitle{A survey of app store analysis for software
  engineering}.
\newblock \bibinfo{journal}{\emph{IEEE transactions on software engineering}}
  \bibinfo{volume}{43}, \bibinfo{number}{9} (\bibinfo{year}{2017}),
  \bibinfo{pages}{817--847}.
\newblock


\bibitem[Miniukovich and De~Angeli(2015)]%
        {miniukovich2015computation}
\bibfield{author}{\bibinfo{person}{Aliaksei Miniukovich} {and}
  \bibinfo{person}{Antonella De~Angeli}.} \bibinfo{year}{2015}\natexlab{}.
\newblock \showarticletitle{Computation of interface aesthetics}. In
  \bibinfo{booktitle}{\emph{Proceedings of the 33rd Annual ACM Conference on
  Human Factors in Computing Systems}}. \bibinfo{pages}{1163--1172}.
\newblock


\bibitem[Ren et~al\mbox{.}(2015)]%
        {ren2015faster}
\bibfield{author}{\bibinfo{person}{Shaoqing Ren}, \bibinfo{person}{Kaiming He},
  \bibinfo{person}{Ross Girshick}, {and} \bibinfo{person}{Jian Sun}.}
  \bibinfo{year}{2015}\natexlab{}.
\newblock \showarticletitle{Faster r-cnn: Towards real-time object detection
  with region proposal networks}.
\newblock \bibinfo{journal}{\emph{Advances in neural information processing
  systems}}  \bibinfo{volume}{28} (\bibinfo{year}{2015}).
\newblock


\bibitem[Rosala(2020)]%
        {nng_user_freedom}
\bibfield{author}{\bibinfo{person}{Maria Rosala}.}
  \bibinfo{year}{2020}\natexlab{}.
\newblock \bibinfo{title}{User Control and Freedom (Usability Heuristic)}.
\newblock
\newblock
\urldef\tempurl%
\url{https://www.nngroup.com/articles/user-control-and-freedom/}
\showURL{%
\tempurl}


\bibitem[Sauer and Sonderegger(2009)]%
        {prototype_fidelity}
\bibfield{author}{\bibinfo{person}{Juergen Sauer} {and}
  \bibinfo{person}{Andreas Sonderegger}.} \bibinfo{year}{2009}\natexlab{}.
\newblock \showarticletitle{The influence of prototype fidelity and aesthetics
  of design in usability tests: Effects on user behaviour, subjective
  evaluation and emotion}.
\newblock \bibinfo{journal}{\emph{Applied ergonomics}} \bibinfo{volume}{40},
  \bibinfo{number}{4} (\bibinfo{year}{2009}), \bibinfo{pages}{670--677}.
\newblock


\bibitem[Shneiderman(2020)]%
        {shneiderman2020human}
\bibfield{author}{\bibinfo{person}{Ben Shneiderman}.}
  \bibinfo{year}{2020}\natexlab{}.
\newblock \showarticletitle{Human-centered artificial intelligence: Reliable,
  safe \& trustworthy}.
\newblock \bibinfo{journal}{\emph{International Journal of Human--Computer
  Interaction}} \bibinfo{volume}{36}, \bibinfo{number}{6}
  (\bibinfo{year}{2020}), \bibinfo{pages}{495--504}.
\newblock


\bibitem[Statista(2022)]%
        {gPlaystoreAppNum}
\bibfield{author}{\bibinfo{person}{Statista}.} \bibinfo{year}{2022}\natexlab{}.
\newblock \bibinfo{title}{Biggest app stores in the world 2022 | Statista.}
\newblock
\newblock
\urldef\tempurl%
\url{https://www.statista.com/statistics/276623/number-of-apps-available-in-leading-app-stores/}
\showURL{%
\tempurl}


\bibitem[Sun et~al\mbox{.}(2021)]%
        {sun2021mind}
\bibfield{author}{\bibinfo{person}{Zhichuang Sun}, \bibinfo{person}{Ruimin
  Sun}, \bibinfo{person}{Long Lu}, {and} \bibinfo{person}{Alan Mislove}.}
  \bibinfo{year}{2021}\natexlab{}.
\newblock \showarticletitle{Mind your weight (s): A large-scale study on
  insufficient machine learning model protection in mobile apps}. In
  \bibinfo{booktitle}{\emph{30th USENIX Security Symposium (USENIX Security
  21)}}. \bibinfo{pages}{1955--1972}.
\newblock


\bibitem[van Berkel et~al\mbox{.}(2021)]%
        {berkel}
\bibfield{author}{\bibinfo{person}{Niels van Berkel}, \bibinfo{person}{Mikael~B
  Skov}, {and} \bibinfo{person}{Jesper Kjeldskov}.}
  \bibinfo{year}{2021}\natexlab{}.
\newblock \showarticletitle{Human-AI interaction: intermittent, continuous, and
  proactive}.
\newblock \bibinfo{journal}{\emph{Interactions}} \bibinfo{volume}{28},
  \bibinfo{number}{6} (\bibinfo{year}{2021}), \bibinfo{pages}{67--71}.
\newblock


\bibitem[Xu et~al\mbox{.}(2019)]%
        {xu2019first}
\bibfield{author}{\bibinfo{person}{Mengwei Xu}, \bibinfo{person}{Jiawei Liu},
  \bibinfo{person}{Yuanqiang Liu}, \bibinfo{person}{Felix~Xiaozhu Lin},
  \bibinfo{person}{Yunxin Liu}, {and} \bibinfo{person}{Xuanzhe Liu}.}
  \bibinfo{year}{2019}\natexlab{}.
\newblock \showarticletitle{A first look at deep learning apps on smartphones}.
  In \bibinfo{booktitle}{\emph{The World Wide Web Conference}}.
  \bibinfo{pages}{2125--2136}.
\newblock


\bibitem[Xu et~al\mbox{.}(2022)]%
        {transitioning}
\bibfield{author}{\bibinfo{person}{Wei Xu}, \bibinfo{person}{Marvin~J Dainoff},
  \bibinfo{person}{Liezhong Ge}, {and} \bibinfo{person}{Zaifeng Gao}.}
  \bibinfo{year}{2022}\natexlab{}.
\newblock \showarticletitle{Transitioning to human interaction with AI systems:
  New challenges and opportunities for HCI professionals to enable
  human-centered AI}.
\newblock \bibinfo{journal}{\emph{International Journal of Human--Computer
  Interaction}} (\bibinfo{year}{2022}), \bibinfo{pages}{1--25}.
\newblock


\bibitem[Yang(2017)]%
        {role_yang}
\bibfield{author}{\bibinfo{person}{Qian Yang}.}
  \bibinfo{year}{2017}\natexlab{}.
\newblock \showarticletitle{The role of design in creating
  machine-learning-enhanced user experience}. In \bibinfo{booktitle}{\emph{2017
  AAAI spring symposium series}}.
\newblock


\bibitem[Yang(2020)]%
        {Profile_AI_Yang}
\bibfield{author}{\bibinfo{person}{Qian Yang}.}
  \bibinfo{year}{2020}\natexlab{}.
\newblock \emph{\bibinfo{title}{Profiling Artificial Intelligence as a Material
  for User Experience Design}}.
\newblock \bibinfo{thesistype}{Ph.\,D. Dissertation}. \bibinfo{school}{Carnegie
  Mellon University}.
\newblock


\bibitem[Yang et~al\mbox{.}(2018a)]%
        {yang2018mapping}
\bibfield{author}{\bibinfo{person}{Qian Yang}, \bibinfo{person}{Nikola
  Banovic}, {and} \bibinfo{person}{John Zimmerman}.}
  \bibinfo{year}{2018}\natexlab{a}.
\newblock \showarticletitle{Mapping machine learning advances from hci research
  to reveal starting places for design innovation}. In
  \bibinfo{booktitle}{\emph{Proceedings of the 2018 CHI conference on human
  factors in computing systems}}. \bibinfo{pages}{1--11}.
\newblock


\bibitem[Yang et~al\mbox{.}(2018b)]%
        {investigating_yang}
\bibfield{author}{\bibinfo{person}{Qian Yang}, \bibinfo{person}{Alex Scuito},
  \bibinfo{person}{John Zimmerman}, \bibinfo{person}{Jodi Forlizzi}, {and}
  \bibinfo{person}{Aaron Steinfeld}.} \bibinfo{year}{2018}\natexlab{b}.
\newblock \showarticletitle{Investigating how experienced UX designers
  effectively work with machine learning}. In
  \bibinfo{booktitle}{\emph{Proceedings of the 2018 Designing Interactive
  Systems Conference}}. \bibinfo{pages}{585--596}.
\newblock


\bibitem[Yang et~al\mbox{.}(2020)]%
        {re_examining}
\bibfield{author}{\bibinfo{person}{Qian Yang}, \bibinfo{person}{Aaron
  Steinfeld}, \bibinfo{person}{Carolyn Ros{\'e}}, {and} \bibinfo{person}{John
  Zimmerman}.} \bibinfo{year}{2020}\natexlab{}.
\newblock \showarticletitle{Re-examining whether, why, and how human-AI
  interaction is uniquely difficult to design}. In
  \bibinfo{booktitle}{\emph{Proceedings of the 2020 chi conference on human
  factors in computing systems}}. \bibinfo{pages}{1--13}.
\newblock


\end{thebibliography}
\end{document}